\documentclass[aps,notitlepage,nofootinbib,longbibliography,twocolumn,superscriptaddress,10pt]{revtex4-1}
\usepackage{amsmath}
\usepackage{amsthm, amssymb}
\usepackage{slashed}
\usepackage{graphicx}
\usepackage{xcolor}
\usepackage{tikz}
\usetikzlibrary{calc,fadings,decorations.pathreplacing,shapes,shapes.multipart,arrows,shapes.misc,intersections,positioning,patterns}
\usepackage{bm}
\usepackage[colorlinks=true,citecolor=blue,linkcolor=magenta]{hyperref}
\usepackage{dsfont}
\usepackage{verbatim}
\usepackage{changepage}
\usepackage{array}
\usepackage{makecell}

\newcommand{\bit}{\begin{itemize}}
\newcommand{\eit}{\end{itemize}}

\renewcommand{\>}{\right\rangle}
\newcommand{\<}{\left\langle}
\newcommand{\ba}{\begin{align}}
\newcommand{\ea}{\end{align}}
\newcommand{\be}{\begin{equation}}
\newcommand{\ee}{\end{equation}}
\newcommand{\bi}{\begin{itemize}}
\newcommand{\ei}{\end{itemize}}

\newcommand{\Tr}{\operatorname{Tr}}

\def\a{\alpha}
\def\b{\beta}
\def\s{\sigma}
\newcommand{\bra}[1]{\< #1 \right|}
\newcommand{\ket}[1]{\left| #1 \>}

\begin{document}
\title{Measurement--Induced  Phase Transitions in the Dynamics of Entanglement}

\author{Brian Skinner}
\affiliation{Department of Physics, Massachusetts Institute of
Technology, Cambridge, MA 02139, USA}
\author{Jonathan Ruhman}
\affiliation{Department of Physics, Massachusetts Institute of
Technology, Cambridge, MA 02139, USA}
\affiliation{Department of Physics, Bar-Ilan University, Ramat Gan 5290002, Israel}
\author{Adam Nahum}
\affiliation{Theoretical Physics, Oxford University, 1 Keble Road, Oxford OX1 3NP, United Kingdom}

\date{\today}

\begin{abstract}

We define dynamical universality classes for many-body systems whose unitary evolution is punctuated by projective measurements.  In cases where such measurements occur randomly at a finite rate $p$ for each degree of freedom, we show that the system has two dynamical phases: `entangling' and `disentangling'.  The former occurs for $p$ smaller than a critical rate $p_c$, and is characterized by volume-law entanglement in the steady-state and `ballistic' entanglement growth after a quench.  
By contrast, for $p > p_c$ the system can sustain only area-law entanglement. At $p = p_c$ the steady state is scale-invariant and, in 1+1D, the entanglement grows logarithmically after a quench.

To obtain a simple heuristic picture for the entangling-disentangling transition, we first construct a toy model that describes the zeroth R\'{e}nyi entropy in discrete time.  We solve this model exactly by mapping it to an optimization problem in classical percolation.

The generic entangling-disentangling transition can be diagnosed using the von Neumann entropy and higher R\'{e}nyi entropies, and it shares many  qualitative features with the toy problem.  We study the generic transition numerically in quantum spin chains, and show that the phenomenology of the two phases is similar to that of the toy model, but with distinct `quantum' critical exponents, which we calculate numerically in $1+1$D.

We examine two different cases for the unitary dynamics: Floquet dynamics for a nonintegrable Ising model, and random circuit dynamics.  We obtain compatible universal properties in each case, indicating that the entangling-disentangling phase transition is generic for projectively measured many-body systems.  We discuss the significance of this transition for numerical calculations of quantum observables in many-body systems.

\end{abstract}

\maketitle

\section{Introduction}

When left unobserved, quantum systems tend to evolve toward states of higher entanglement
\cite{calabrese2005evolution,kim2013ballistic,liu2014entanglement,kaufman2016quantum,asplund2015entanglement,casini2016spread,hosur_chaos_2016,ho2017entanglement,nahum_quantum_2017,mezei2017entanglement,mezei2017entanglement2,gu_spread_2017,jonay_coarse-grained_2018,mezei2018membrane,de2006entanglement, vznidarivc2008many,
bardarson2012unbounded,zhou2018emergent}.  
Unitary evolution of a many-body wavefunction, with a Hamiltonian or with quantum gates,
typically drives it towards a state with volume-law scaling for the entanglement entropies of subsystems.
By contrast, local measurements can reduce the entanglement in a quantum system by collapsing degrees of freedom.  A measurement of a single spin--1/2, say, leaves that spin in a definite spin state, and disentangles it from the rest of the system.

What happens to the entanglement in a quantum system when measurements occur repeatedly during the evolution at a fixed rate?  For simplicity, let us model the  local measurements as occurring at random times and locations, at a nonzero rate $p$ per degree of freedom.
Do such measurements collapse the many-body wavefunction into something close to a product state, with area law entanglement, or can volume law entanglement survive?

\begin{figure}[b]
 \begin{center}
 \includegraphics[width=0.75\linewidth]{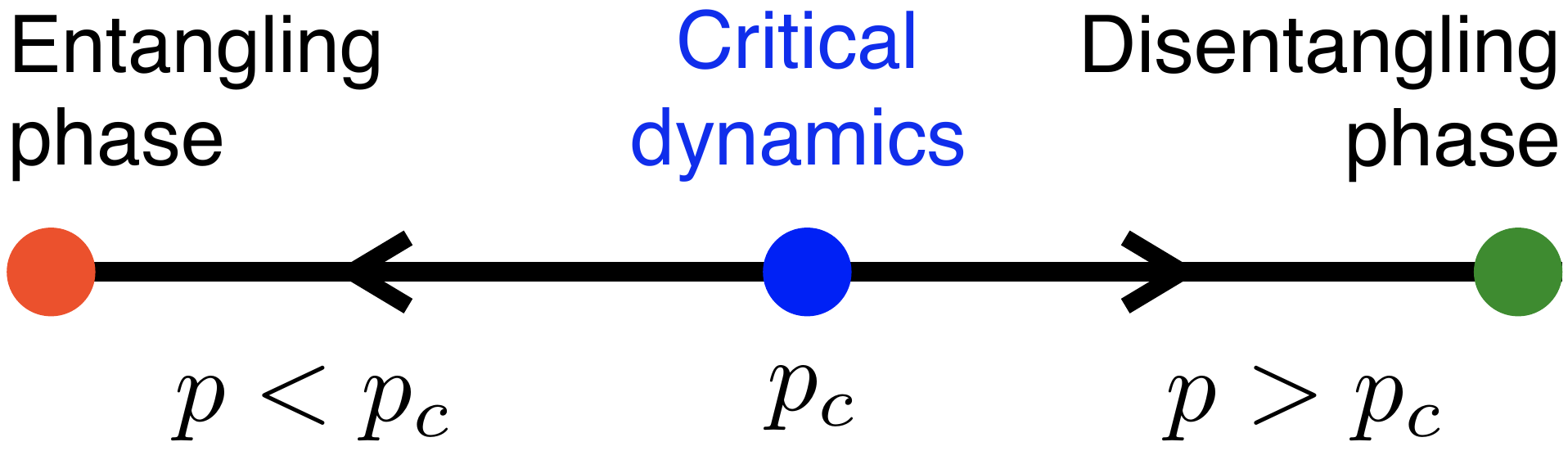}
 \end{center}
\caption{Phase diagram as a function of $p$, the rate at which measurements are made for each degree of freedom. Arrows indicate renormalization group flow.}
\label{fig:phasediagram}
\end{figure}

Here we answer this question for the simplest type of measurement, which is a projective measurement of a discrete degree of freedom. We show that both types of dynamics can occur, leading to volume-law or area-law entanglement, depending on the value of the measurement rate $p$. These two distinct dynamical phases, `entangling' and `disentangling', are separated by a continuous phase transition at a finite value of $p = p_c$ (Fig.~\ref{fig:phasediagram}).  (We call the latter phase `disentangling' because if a finite system is initiated in a volume-law entangled state, the dynamics will eventually reduce the entanglement to area law.)

In this paper we introduce two versions  of this transition:
 the `generic' entanglement phase transition and a toy model for it that is amenable to exact analysis.
We show in later parts of the paper that the `generic' transition occurs in models that are not fine-tuned, including physically sensible ones, and that it affects the dynamics of the von Neumann entanglement entropy and of certain equal-time correlation functions.
First, however, we address the toy model for the transition in depth. While the toy model is in a different universality class from the generic transition, it
captures many qualitative features of the phase diagram and the transition remarkably well.

The toy model is an exact description of the dynamics of the \textit{zeroth} R\'{e}nyi entanglement entropy ($S_0$) in a system with discrete-time dynamics that has a circuit representation.\footnote{The zeroth R\'{e}nyi entropy (Hartley entropy) is perhaps most familiar in 1D, where it is the logarithm of the bond dimension needed for an exact matrix product state representation of the system.}

For circuit  dynamics without measurement, the well-known `minimal cut' formula gives $S_0$ exactly as a function of time so long as the dynamics is not fine-tuned (see Ref.\ \onlinecite{nahum_quantum_2017} for a rigorous proof in one setting).  We show that the minimal cut formula still holds exactly when there are projective measurements --- our key insight is that it should be applied to a network in which some bonds are `broken' by the presence of a measurement.

\begin{figure}[t]
 \begin{center}
 \includegraphics[width=0.9\linewidth]{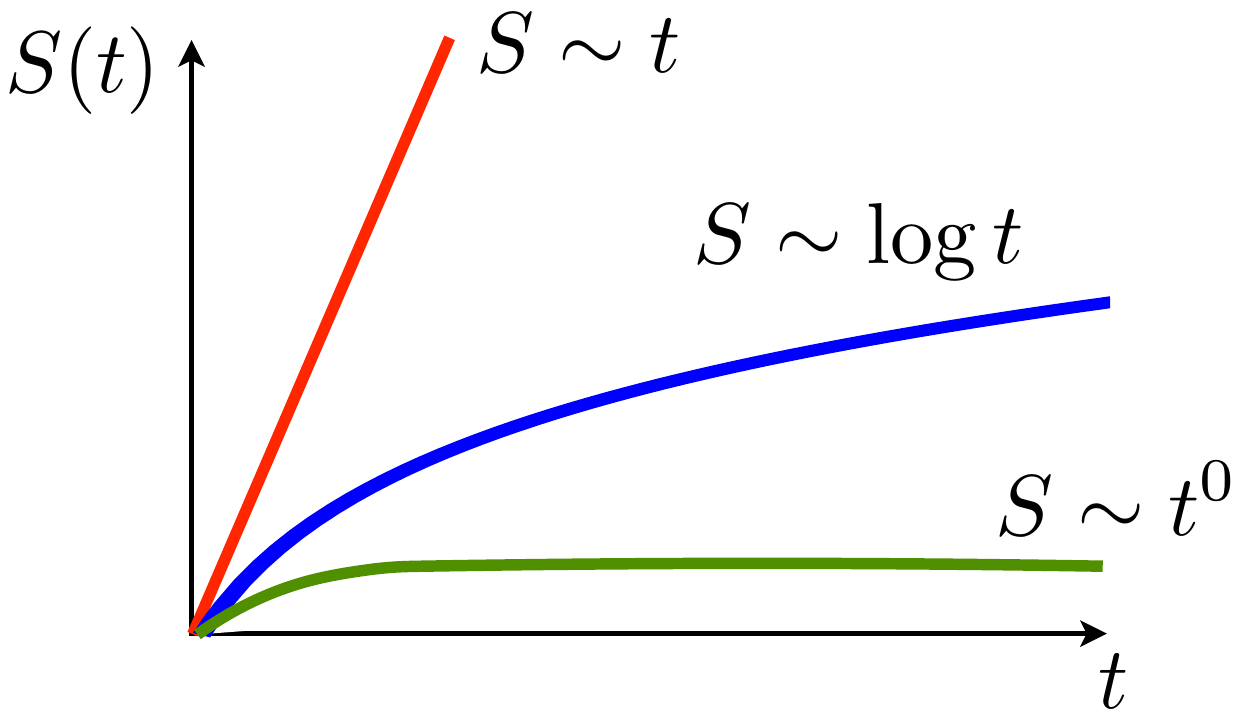}
 \end{center}
\caption{Schematic illustration of entanglement production after a quench from a product state in 1+1D.
The growth of bipartite entanglement entropy between the two semi-infinite halves of an infinite chain is shown.
In the entangling phase ($p<p_c$; upper curve) the entanglement grows `ballistically' with time.
At the critical point ($p=p_c$; middle curve), the entanglement grows logarithmically.
In the disentangling phase ($p>p_c$; lower curve), the entanglement saturates to a finite value. (Random fluctuations are averaged over.)
}
\label{fig:growthschematic}
\end{figure}

The minimal cut representation of $S_0$ yields an effective classical optimization problem: finding the optimal cut through a bond percolation configuration. We use this mapping to characterize the scaling behavior of the entanglement ($S_0$) and mutual information ($I_0$) at and on both sides of the critical point of the toy model, which is at the bond percolation threshold $p_c$. The growth of entanglement in the three regimes is illustrated schematically in Fig.~\ref{fig:growthschematic} for a 1+1D system initialized in an area-law state. \
The logarithmic entanglement growth at $p_c$ is a consequence of scale invariance, which also leads to power law correlations of a certain type between distant spins.
We will show that the three types of growth in Fig.~\ref{fig:growthschematic} also characterize the regimes of the generic problem.

The existence of a transition in the toy model has a very simple interpretation which applies in any spatial dimension $d$. Namely, when the measurement rate exceeds $p_c$,  we effectively break enough bonds for the circuit to fall apart into disconnected pieces.  Such pieces are disentangled from each other, and the circuit no longer mediates long-range correlations.  

As we show below, the `generic' transition occurs at a value of $p_c$ that is smaller than the value suggested by the dynamics of $S_0$.
That is, as $p$ is increased,
entanglement production ceases well \textit{before} the circuit falls apart in the above sense.
We diagnose the generic transition using the von Neumann entanglement entropy (and higher R\'{e}nyi entropies).
We focus on 1+1D spin chains, where quantum simulations are feasible up to at least $L = 24$ using matrix product states  \cite{ITensor}.
The results from the toy model guide our analysis of the data from these systems.
Strikingly, many qualitative features of the toy model continue to hold, and we show clear evidence for a transition at a finite $p_c$.
The universality class of the transition, however, is distinct from classical percolation, as we show by computing the correlation length exponent close to the transition.
In particular, as the transition is approached, the characteristic length and time scales diverge as $\xi \sim |p-p_c|^{-\nu}$ and $\tau\sim |p-p_c|^{-\nu z}$ respectively, with $\nu = 2.03(5)$ and a dynamical exponent $z$ that is consistent with $z = 1$. We obtain consistent exponent estimates for two different models, including a deterministic Floquet circuit and a random unitary circuit (each with random measurements).

The specific models we study all have  discrete time dynamics. While this discretization is important in order for the dynamics of $S_0$ to be well-defined\footnote{For dynamics in continuous time $S_0$ typically becomes infinite as soon as $t>0$.} (i.e. for the construction of the toy model),  we do not expect that it will affect the existence or universality class of the `generic' transition that is manifest in physically meaningful quantities (such as the von Neumann entanglement entropy $S_1$ or the mutual information between separated spins). It is also possible to consider a continuous quantum measurement process, which is obtained as a limiting case of very frequent `weak' measurements \cite{davies1976quantum}.\footnote{In a weak measurement, a probe spin (say) first interacts with the system for a short time, after which its state is measured. This leaves the system in a pure state that is close to its original state if the interaction with the probe was brief.} The production of entanglement in this setting has been considered for free fermions in Ref.~\onlinecite{cao2018entanglement}, where arbitrarily weak measurement was found to lead to an area-law state. This was explained in terms of a quasiparticle picture making use of integrability \cite{cao2018entanglement}. We conjecture that nonintegrable models subjected to continuous measurement behave similarly to the models we study here.

 We briefly discuss the outlook for an analytical description of the dynamical transition that we find, making connections with recent results for random unitary circuits
\cite{nahum_quantum_2017,chan_solution_2017,jonay2018coarse,zhou2018emergent}  and random tensor network states \cite{hayden2016holographic,vasseur2018entanglement}. 
 Recently, Ref.~\onlinecite{vasseur2018entanglement} discussed an entanglement phase transition in a random tensor network state \cite{hayden2016holographic} as the bond dimension was varied. We discuss the possibility that the conformal field theory discussed there also describes the dynamical transition.

The entanglement structure of a quantum state has a direct bearing on how difficult it is to simulate the dynamics of that state using a classical computer \cite{verstraete2006matrix}. Thus the entanglement transition may have important implications for simulating quantum evolution with measurement, as we discuss in Sec.\ \ref{sec:implications}.

Finally, we mention earlier work by Aharonov \cite{Aharonov}, who considered the possibility of a transition in which the state of a quantum computer is affected by  decoherence. Aharonov used the percolation connectivity of the circuit to demonstrate that when decoherence events are sufficiently frequent the \textit{mixed state} of the qubits must be associated with a finite entanglement length scale. 
On the other hand, in the presence of active quantum error correction, a very low rate of decoherence can allow for nontrivial entanglement of qubits implementing a certain algorithm.
We emphasize, however, that the transition envisaged in Ref.\ \onlinecite{Aharonov} is in the density matrix of a \emph{mixed} state (with the mixing arising from environmental decoherence). In the problem we study here, on the other hand, there is no transition of this type: the mixed state obtained by averaging over measurements is trivial throughout the phase diagram, as we emphasize in Sec.\ \ref{sec:models}. 
The transition we introduce is in the entanglement structure of pure states.

\tableofcontents

\section{Models and setting}
\label{sec:models}

The dynamics we study consist of unitary evolution of a {spin-1/2} chain interspersed with single-spin  measurements.
We use quantum dynamics in discrete time, where each time step involves the application of unitary gates to pairs of adjacent spins.  This allows us to define a nontrivial solvable toy model (the discrete-time dynamics of $S_0$) and it simplifies the numerical study of the `generic' problem.
The dynamical protocol we describe could be modified in various ways, but we expect the universality class of our `generic' transition to be robust (see the discussion at the end of this section).

\begin{figure}[t]
 \begin{center}
 \includegraphics[width=0.6\linewidth]{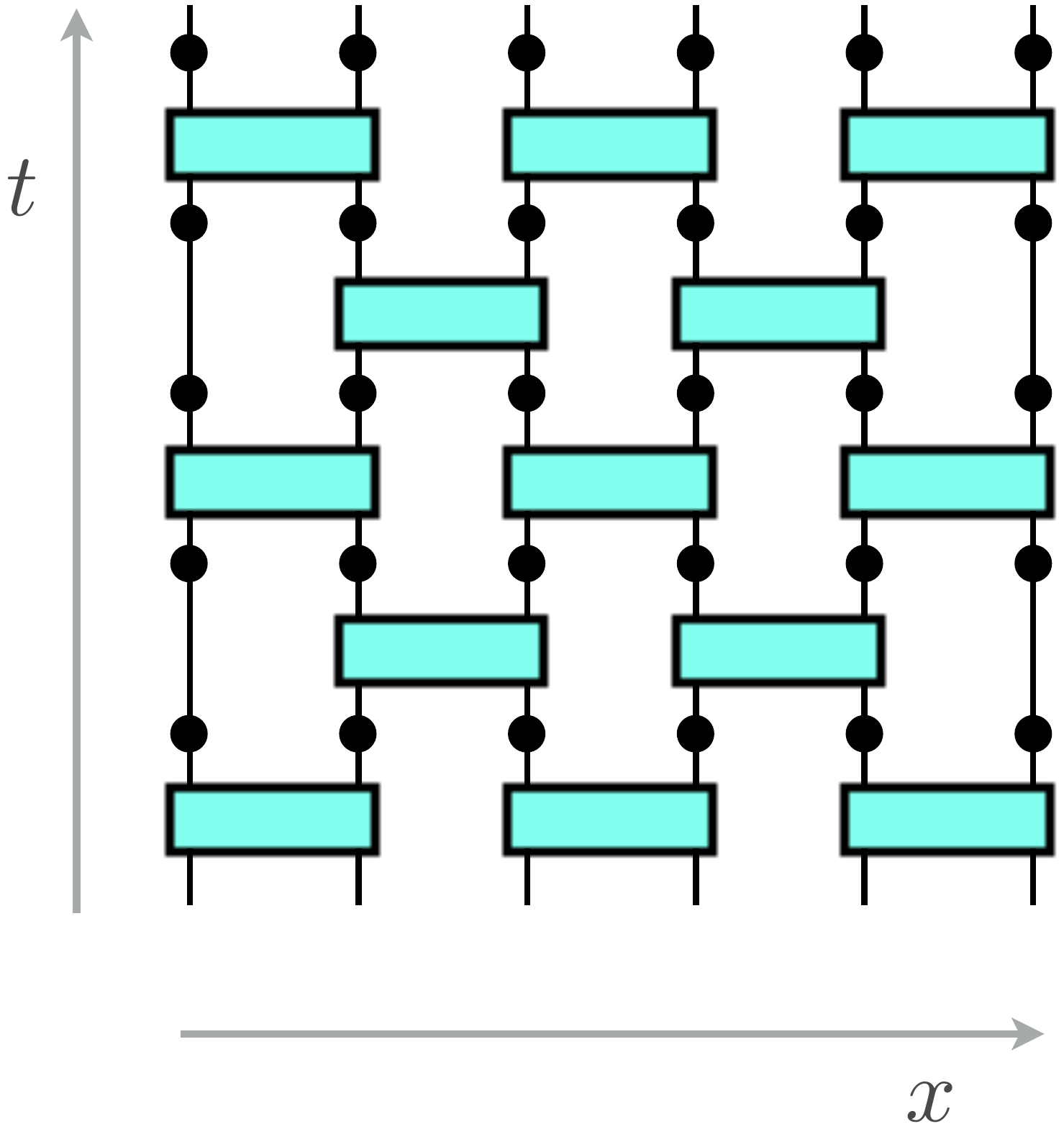}
 \end{center}
\caption{ Circuit representation for evolution of the quantum system.
Bricks indicate unitary operators (specified in the text).
Dots indicate spacetime locations where measurements may take place.
Note that the full dynamics is \textit{nonlinear}, since the state must be re-normalized after each projective measurement event. 
}
\label{fig:firstcircuitfig}
\end{figure}

Our 1+1D quantum circuits are arranged with a `running bond' configuration of unitaries, as in Fig.\ \ref{fig:firstcircuitfig}.
We define units of time such that one time step involves applying one layer of unitaries (the time period of the circuit is two layers).

Measurement events take place randomly: after each layer of unitaries, each spin has a probability $p$ of having its $z$ component ($S_z$) measured. 
A detail regarding the boundary conditions is that Fig.\ \ref{fig:firstcircuitfig} shows a layout  in which the two boundary spins have two opportunities to be measured for each unitary that is applied to them: we use this layout in Sec.~\ref{sec:zerothRenyi} as it is  natural for our classical mapping, but for the quantum simulations in Sec.~\ref{sec:genericdynamicaltransition} the boundary spins have only one chance to be measured per unitary applied to them.

In addition to the randomness in the times and locations of measurements, which we have put in by hand, there is intrinsic randomness in the measurement \textit{outcomes}, which occur with the usual probabilities ${|\langle \uparrow | \Psi(t) \rangle|^2}$ and ${|\langle \downarrow | \Psi (t) \rangle|^2}$, where $\ket{\Psi(t)}$ is the state prior to the measurement. 
After measuring we project onto the value of $S_z$ obtained. The state must be re--normalized after each projective measurement, so its dynamics are both stochastic and nonlinear. Nonetheless, the state remains pure. This pure state determines the probabilities for measurements, conditioned on the outcomes of previous measurements.\footnote{It should be borne in mind that conventional expectation values, if we \textit{average} them over all measurement outcomes, lose the information about the entanglement structure of this state (see the discussion towards the end of this section).}

We simulate two types of quantum dynamics:

(i) Random unitary dynamics, in which every `brick' in Fig.~\ref{fig:firstcircuitfig} is chosen randomly and independently from the unitary group (with the Haar measure). Studying  1+1D random unitary circuits and related models has elucidated the coarse-grained dynamics of entanglement and quantum operators in more realistic systems \cite{nahum_quantum_2017,jonay_coarse-grained_2018, zhou2018emergent, nahum_operator_2017, von_keyserlingk_operator_2017, khemani_velocity-dependent_2018, rakovszky2017diffusive, khemani_operator_2017, chan_solution_2017, chan2018spectral,kos2017many, dahlsten2007emergence,vznidarivc2008exact, rowlands2018noisy, knap2018entanglement, AnushyaChris}.
Here we extend the model to dynamics with measurement.

(ii) `Floquet' dynamics, in which the unitary circuit is deterministic and periodic in time (in the absence of measurements). We study this case in order to check that our results are not dependent on the randomness of the unitaries in case (i). The fixed unitary we use (the elementary brick for Fig.~\ref{fig:firstcircuitfig}) is a product of two non-commuting unitaries. For spins at positions $x$ and $x+1$ the unitary is ($\hat S_i = \hat \sigma_i /2$):
\ba\label{eq:unitarydefinition}
U_{x,x+1} &  = e^{ - i \left[ \phi_{xx} \hat \sigma_x(x) \hat \sigma_x(x+1) +  \theta_x   \hat \sigma_x(x)+  \theta_x   \hat \sigma_x(x+1)  \right] } \nonumber \\
& \times  e^{ - i \left[ \phi_{zz} \hat \sigma_z(x) \hat \sigma_z(x+1) + \theta_z   \hat \sigma_z(x)+ \theta_z   \hat \sigma_z(x+1)  \right]}\,.
\end{align}
where $\phi_{xx} = 0.3$, $\phi_{zz} = 0.6$, $\theta_x = 0.2$ and $\theta_z =0.4 $.
This unitary operation defines a version of the Floquet Ising model with longitudinal and transverse fields.

Our main tools for characterizing the dynamics are the R\'{e}nyi entanglement entropies for subsystems $A$ of the spin system,
\be
S_n(A) = \frac{1}{1-n} \log_2 \Tr_A \rho_A^{n},
\label{eq:Sndef}
\ee
where $\rho_A = \Tr_{\overline{A}} \ket{\Psi}\bra{\Psi}$ is the reduced density matrix of  $A$.  At $n \rightarrow 1$, this definition reproduces the von Neumann entropy
\be
S_1 (A) = - \Tr_A \rho_A \log_2 \rho_A.
\ee
We measure all entropies in bits.

It is worthwhile to note that the entanglement entropy $S_n$ for $n>1$ is constrained by the inequalities \cite{Wilming2018entanglement}
\be 
S_\infty \leq S_n \leq \frac{n}{n-1} S_\infty,
\ee 
where $S_\infty = \lim_{n\rightarrow \infty} S_n = \log_2(1/\lambda_\textrm{max})$, and $\lambda_\textrm{max}$ is the largest eigenvalue of $\rho_{A}$.
These inequalities imply that any entanglement entropy $S_n$ with $n > 1$ can differ from $S_\infty$ by at most a constant factor $n/(n-1)$, and consequently all $S_n$ with $n > 1$ must have the same scaling with system size.  As we show below, this is born out in our numerical results, which indicate a single transition for all $S_n$ with $n \geq 1$.

We also study the R\'{e}nyi mutual information $I_n ({\bf a} , {\bf b})$ between two separated spins (labelled here ${\bf a}$  and ${\bf b}$):
\be\label{eq:I0defn}
I_n ({\bf a} , {\bf b})\equiv S_n({\bf{a}}) + S_n({\bf{b}}) - S_n({\bf{a}} \cup {\bf{b}}). \nonumber
\ee
In a random system the $I_n$ are useful measures of the strength of quantum correlations between spins ${\bf{a}}$ and ${\bf{b}}$, because they are independent of the choice of local bases for these spins.\footnote{$I_2$ is simply related to the natural correlation function ${C = \sum_{ij} \big[
\bra{\Psi} \hat \sigma_i({\bf a}) \hat \sigma_j({\bf b})\ket{\Psi}^2
-
\bra{\Psi} \hat \sigma_i({\bf a})\ket{\Psi}^2
\bra{\Psi}  \hat \sigma_j({\bf b})\ket{\Psi}^2 \big]}$. If $C_0 \equiv \big( 1+ \sum_i \bra{\Psi} \hat \sigma_i({\bf a}) \ket{\Psi}^2\big)  \big( 1+ \sum_i \bra{\Psi} \hat \sigma_i({\bf b}) \ket{\Psi}^2 \big)$ then ${I_2 = \log (1+C/C_0)}$.    $C_0$ is of order 1, so if $C$  tends to zero at large separation of the spins, then $I_2$ tends to zero in the same manner.}
We show below that  in the present case they reveal a remarkable scale-invariant entanglement structure at the transition.

Let us briefly clarify what the dynamical transition does and does not imply.
For simplicity, assume for a moment that the dynamics is deterministic except for the intrinsic randomness in the measurement outcomes.
As noted above, the evolving quantum state
remains pure. This pure state, which we temporarily  denote $\ket{\Psi^{(o_1, \ldots, o_N)}}$,
depends on the previous measurement outcomes $o_1, \ldots, o_N$, and it determines the probabilities for measurement outcomes at a particular time \textit{given} these previous  outcomes.
This pure state must be contrasted with the mixed state that is obtained if we \textit{average} over measurement outcomes. Let us denote the average over measurement outcomes by $\mathbb{E}_{o_1,\ldots}$, and let $\< \cdots \>_{o_1,\ldots}$ be the expectation value of an observable in the state $\ket{\Psi^{(o_1, \ldots, o_N)}}$. The averaged density matrix is 
\be
\rho_\text{av}=
\mathbb{E}_{o_1,\ldots,o_N}  \big|{\Psi^{(o_1, \ldots, o_N)}}\big\rangle \big\langle{\Psi^{(o_1, \ldots, o_N)}}\big|.
\ee
This  mixed state $\rho_\text{av}$ evolves linearly, in a standard way. Generically we expect it simply to evolve to a trivial infinite temperature Gibbs state, for any nonzero rate of measurement. The dynamical phase transition we discuss is therefore \textit{not} apparent in this object, and therefore it is not apparent in correlation functions such as
\be
\mathbb{E}_{o_1,\ldots} \<\hat \sigma^z({\bf a}) \hat \sigma^z({\bf b}) \>_{o_1,\ldots}
\equiv \Tr \rho_\text{av}\hat \sigma^z({\bf a}) \hat \sigma^z({\bf b})
\ee
that are straightforwardly averaged over previous measurement outcomes. 
As noted above, the transition is detected by more complex local correlation functions, for example
\be
\mathbb{E}_{o_1,\ldots} 
\left[
\<\hat \sigma^z({\bf a}) \hat \sigma^z({\bf b}) \>_{o_1,\ldots} ^2 -
\<\hat \sigma^z({\bf a}) \>_{o_1,\ldots} ^2 \<\hat \sigma^z({\bf b}) \>_{o_1,\ldots} ^2 
\right].
\ee
The squares inside the average imply that experimental measurement of such a quantity would be a severe statistical challenge, as a result of the need for extensive postselection. Therefore a more likely application of our results is to questions of computational difficulty (see Sec.~\ref{sec:implications}).

 \begin{table}[t]
\begin{center}
    \begin{tabular}{ | c | c | c | c |}
    \hline
    \thead{R\'{e}nyi \\ entropy \\ order, $n$} & \thead{simulation \\ dynamics} &   \thead{critical \\ measurement \\ rate, $p_c$} & \thead{correlation \\ length \\ exponent $\nu$} \\ \hline\hline
& (exact) & $1/2$ & $4/3$ \\ \cline{2-4}
      & \makecell{classical \\ simulation} & $0.51 \pm 0.01$ & $1.24 \pm 0.13$ \\ \cline{2-4}
    0  & \makecell{random \\  unitaries} & $0.50 \pm 0.05$ & $1.36 \pm 0.10$ \\ \cline{2-4}
      & \makecell{Floquet \\ dynamics} & $0.51 \pm 0.01$ & $1.36 \pm 0.21$ \\ \hline\hline

    1 & \makecell{ } & $0.26 \pm 0.08$ & $2.01 \pm 0.10$ \\ \cline{1-1} \cline{3-4}
    2 & \makecell{random \\ unitaries} & $0.27 \pm 0.03$ & $2.31 \pm 0.75$ \\ \cline{1-1} \cline{3-4}
   $\infty$ & \makecell{ } & $0.26 \pm 0.08$ &  $2.25 \pm 0.25$  \\ \hline \hline

     1 & \makecell{} & $0.21 \pm 0.16$ & $1.74 \pm 0.45$ \\ \cline{1-1} \cline{3-4} 
     2  & \makecell{Floquet  \\ dynamics} & $0.22 \pm 0.05$ & $2.07 \pm 0.22$  \\ \cline{1-1} \cline{3-4} 
      $\infty$  & \makecell{} & $0.21 \pm 0.05$ & $2.03 \pm 0.07$ \\ \hline
    \end{tabular}
    \caption{Critical measurement rate $p_c$ and correlation length exponent $\nu$ obtained numerically by studying  different orders $n$ of the R\'{e}nyi entropy $S_n$, 
    and different dynamical protocols.
}
    \label{tab:exps}
\end{center}
\end{table}

To end this section, let us discuss the issue of robustness of our results to the choice of protocol.
Recall that we study both a toy model that describes the transition of $S_0$, and the more physical (generic) transition that is captured by the higher R\'{e}nyi entropies. The question of robustness is mostly of interest for the latter, but let us first comment on the former.

The dynamics of $S_0$ (specifically) is very sensitive to the fact that the dynamics is of circuit type. $S_0$ is not typically a physically significant quantity\footnote{Despite this, for certain types of unitary dynamics $S_0$ coincides with the von Neumann entropy. This includes dynamics generated by unitary gates in the Clifford group \cite{gottesman1998heisenberg}, and random unitary circuits in the limit of infinite local Hilbert space dimension.} because of its extreme sensitivity to weak perturbations of the state. Nevertheless, within the setting of circuit dynamics, the universal results we find for $S_0$ are robust for that quantity. Our classical percolation mapping is exact for any circuit of the form shown in Fig.~\ref{fig:firstcircuitfig}, \textit{irrespective} of the choice of unitaries in the circuit, so long as they are not fine-tuned.

While a full presentation of results is left for later sections, we note here that, in terms of the physical transition, our results suggest that protocols (i) and (ii) listed above are in the same universality class. In Table~\ref{tab:exps} we show the location of the critical point (which, of course, depends on the nature of the unitary operator being applied and is therefore nonuniversal), and the correlation length exponent\footnote{The correlation length exponent quantifies the divergence of the characteristic length and time scales as the critical point at $p_c$ is approached; see the following section.} obtained for each type of dynamics by studying the different R\'{e}nyi entropies.  For a given type of dynamics, different R\'{e}nyi entropies with $n\geq 1$ all give compatible estimates of $p_c$, consistent with the idea that there is just a single physical transition in the entanglement structure.

The  protocols we defined above could be varied in many ways. For example, one could replace circuit dynamics with Hamiltonian dynamics in continuous time (with measurements remaining discrete events). We expect this difference to be unimportant for the long-lengthscale physics of the transition.
Note, for example, that there is generically no difference in symmetry between the two situations. In the absence of measurements, continuous time Hamiltonian dynamics (with a constant Hamiltonian) have time translation symmetry, but this symmetry is destroyed
by measurements, which inevitably induce randomness.

The intrinsic randomness of measurement {outcomes} also leads us to conjecture that  the universality class of the transition can persist for dynamics in which the randomness in the \textit{times} and \textit{locations} of measurements is removed, since the dynamics remains random anyway.  For example, for continuous time dynamics, one could make measurements in some periodic fashion, with a variable period $t_0$, and we expect a transition as a function of the dimensionless rate $\hbar /(t_0 J)$, where $J$ is the energy scale of the Hamiltonian.  `Weak' measurements can also be substituted for projective measurements \cite{cao2018entanglement}.

For the rest of the paper we focus on the circuit dynamics (i) and (ii) and higher-dimensional generalizations.

\section{Zeroth R\'{e}nyi entropy and classical optimization}
\label{sec:zerothRenyi}

The zeroth R\'{e}nyi entropy of entanglement between a subsystem and its exterior, $S_0$,
counts the number $N$ of nonzero eigenvalues of the reduced density matrix: $S_0 = \log_2 N$.
Since arbitrarily small eigenvalues are counted, $S_0$ can change discontinuously in response to a small perturbation,
and as a result we do not usually think of it as a physically significant quantity.
Still, in an idealized system it can be a useful conceptual starting point for thinking about the von Neumann and higher-order entanglement entropies.\footnote{It is also an upper bound on these more physical entropies. This bound is saturated in certain limits; see footnote 6.}

The tool for calculating $S_0$ is the `minimal cut', a geometrical upper bound on entanglement in a tensor network which is a useful starting point for  thinking about the scaling of entanglement in various situations ranging from holography \cite{swingle2012entanglement, pastawski2015holographic, hayden2016holographic} to unitary dynamics
\cite{casini2016spread, nahum_quantum_2017,jonay_coarse-grained_2018}.
The minimal cut gives $S_0$ exactly for our dynamics (for both choices of unitaries in Sec.~\ref{sec:models}),  as would be expected from parameter-counting and as we confirm numerically below.\footnote{The minimal cut formula for $S_0$ is known to hold exactly with probability 1 for a 1+1D random unitary circuit applied to an initial product state (i.e. at $p=0$) \cite{nahum_quantum_2017}.
More generally, care must be taken that the unitaries being applied are not fine-tuned.  Omitting the  $\hat \sigma_x\hat \sigma_x$ term from Eq.~(\ref{eq:unitarydefinition}), for example, results in a unitary that can generate a smaller growth in $S_0$ than a generic 2-spin unitary.}

We expect the mapping below for $S_0$ to be exact for the present circuit geometry  for any non-fine-tuned choice of the unitaries in the network, whether random or deterministic.
This mapping is likely also to describe higher R\'{e}nyi entropies in special limiting cases, as discussed in Sec.~\ref{sec:universalityclasses}, though not generically.

\begin{figure}[t]
\begin{center}
\includegraphics[width=\linewidth]{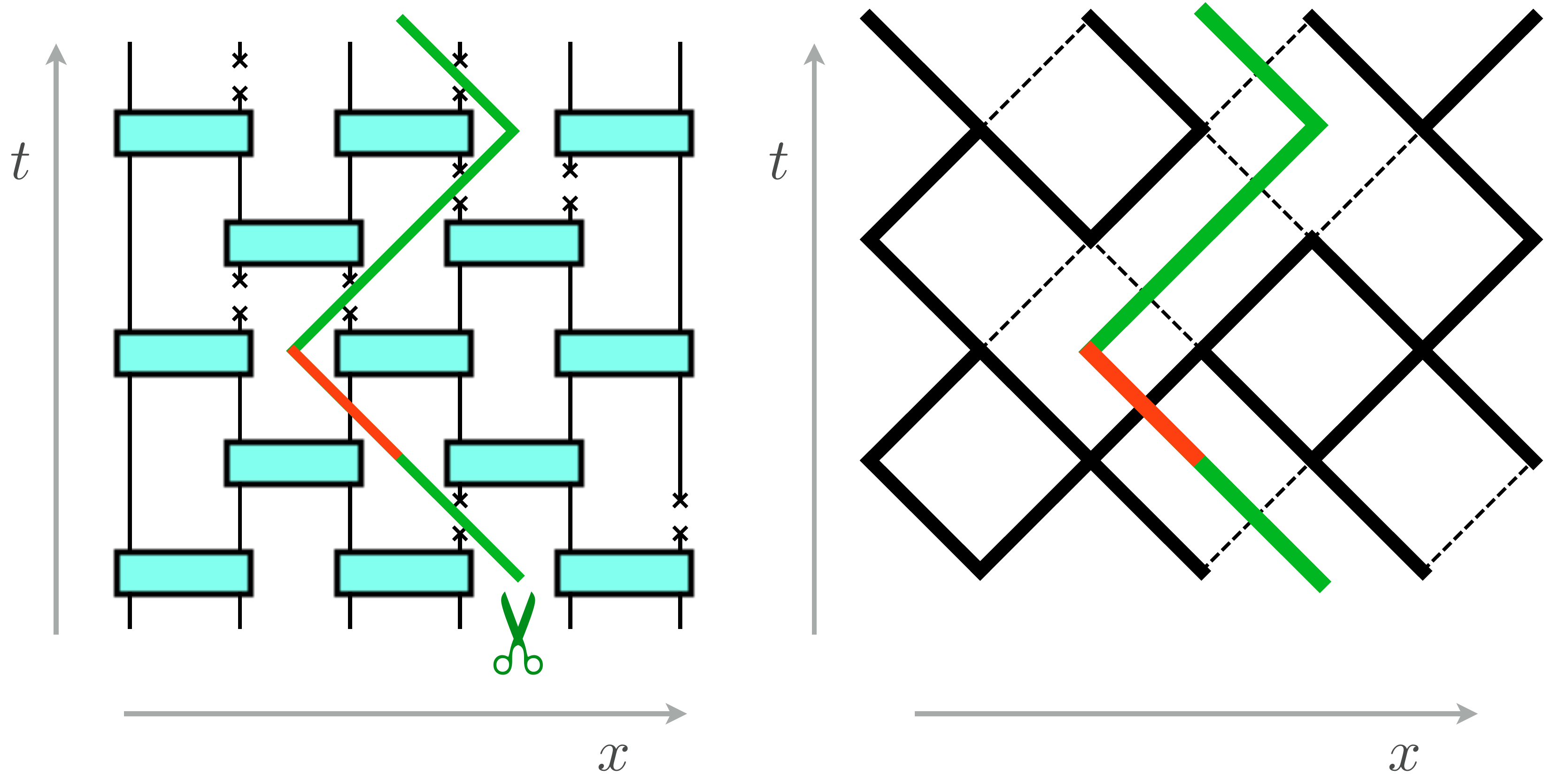}
\end{center}
\caption{Mapping between circuit dynamics with measurement (Left) and bond percolation on the square lattice (Right).
For the purposes of the minimal cut picture, a projective measurement breaks a bond.
The minimal cut may pass through broken bonds at zero cost.
For the figure on the right,
which is topologically equivalent,
we represent each unitary as the vertex of a square lattice.
Broken bonds are dashed (unoccupied) and unbroken bonds are solid (occupied).
This is bond percolation on the square lattice, with a probability $1-p$ for a bond to be occupied.
The minimal cut lives on the dual square lattice.
}
\label{fig:perclatticemapping}
\end{figure}

In the absence of projective measurements, $S_0(A)$ is given exactly by the number of links in the circuit that must be cut in order to separate
the physical `legs' (external bonds at the top boundary) for the spins in $A$ from those for the spins outside $A$.
In the presence of projective measurements, there is a simple generalization of this picture, which is illustrated in Fig.\ \ref{fig:perclatticemapping}.

 When a projective measurement is made, the state of the measured spin at that point in its history is fixed, for example to $\uparrow$. To express the wavefunction (for a given outcome of the measurement) it is then no longer necessary to sum over the spin index on this site. One can therefore think of a measurement as ``breaking a bond'' in the classical network.

 If the broken bonds are sufficiently dense that one subsystem is completely isolated from the other --- i.e., if one can make a cut that separates the two subsystems without passing through any unbroken bonds --- then the two subsystems can have no entanglement.  If no such cut is possible, then the zeroth R\'{e}nyi entropy is equal to the number of unbroken bonds that must be cut in order to separate the two subsystems.

The above picture applies for each set of measurement outcomes, which we should in principle average over using Born's rule. However this picture shows that $S_0$ is independent of the measurement outcomes, and depends only on their times and locations.

In this way the problem of calculating $S_0$ for some subsystem becomes equivalent to an optimization problem in a classical percolation configuration.
Some proportion $p$ of bonds in a network are broken, and one must determine the minimal number of additional bonds that should be cut in order to separate the subsystem from the rest of the network. This is illustrated in Fig.\ \ref{fig:perclatticemapping}.

One can generalize this mapping to any number of dimensions. If the number of spatial dimensions is $d$, the minimal cut is a surface of $d-1$ dimensions (see Sec.~\ref{sec:higherdimensions}). For the rest of this section we restrict our attention to 1+1D, where the minimal cut is a path. Problems of finding a minimal-cost path in a disordered medium are well-studied in the mathematical literature under the name `first passage percolation' \cite{kesten1986aspects}. Reference \onlinecite{chayes1986critical} contains rigorous results for the present case, where steps of the path cost either zero or unity. Below we give a heuristic discussion  that is consistent with Ref.\ \onlinecite{chayes1986critical}.

\begin{figure}[t]
\begin{center}
\includegraphics[width=\linewidth]{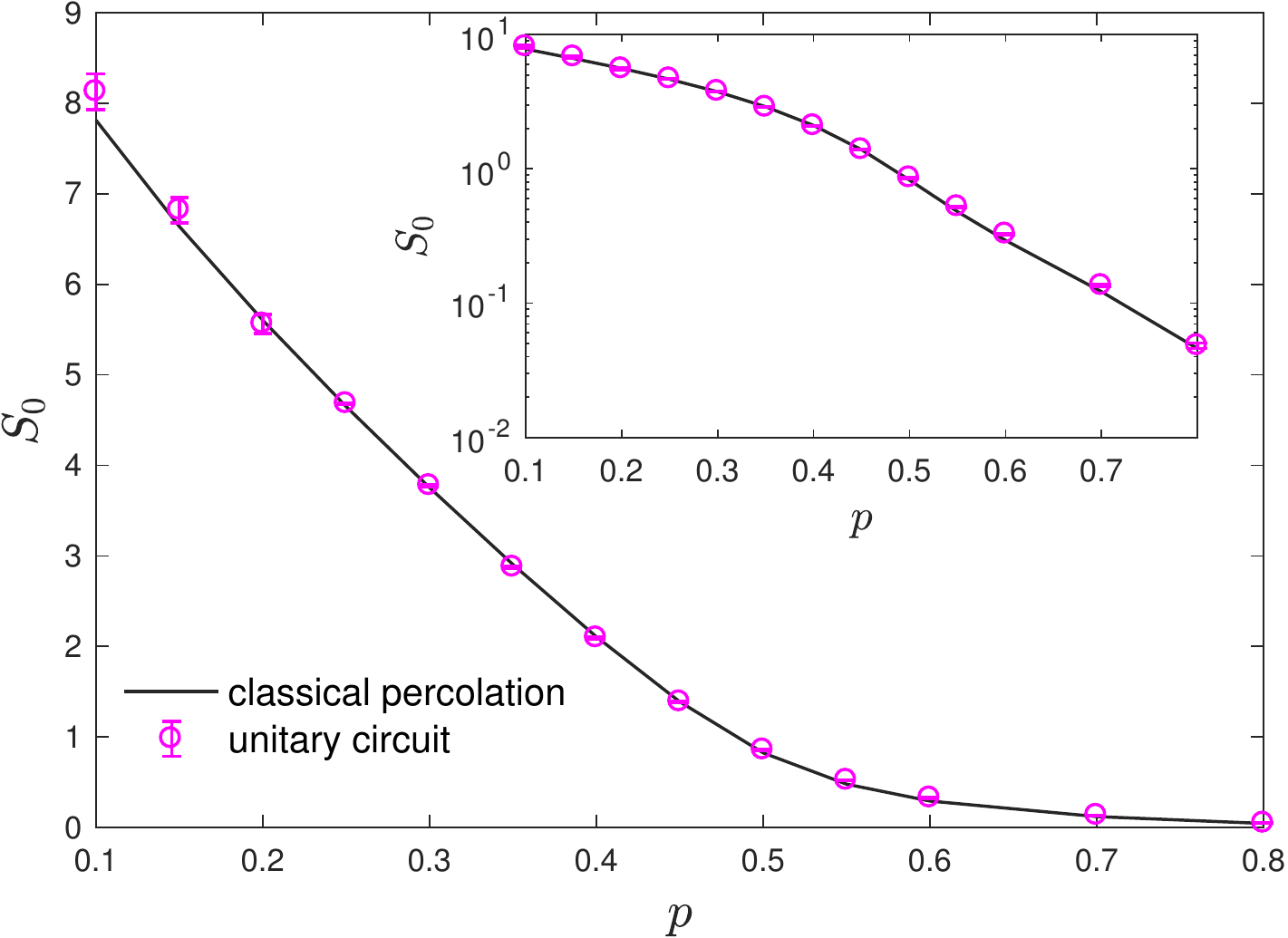}
\end{center}
\caption{Example comparison between the unitary circuit and classical percolation results for the zeroth R\'{e}nyi entropy, $S_0$, between two halves of a spin chain that has undergone unitary evolution.  In this example, the circuit contains $L = 24$ spins and the evolution time is $t = 48$.  The value of $p$ on the horizontal axis represents the probability of measurement for each spin after each time step.  The magenta circles with error bars show the result of the full simulation of the unitary circuit, while the black line shows the result of the classical percolation simulation. The inset shows the same data on a logarithmic scale.}
\label{fig:validation}
\end{figure}

The classical percolation problem can be approached computationally using standard numerical algorithms, which we describe in Appendix~\ref{sec:numericalmethods}.  Briefly, we simulate a rotated square lattice (as shown on the right-hand side of Fig.\ \ref{fig:perclatticemapping}) with width $L$ and depth $t$, of which each bond has a cost of either zero (with probability $p$) or unity (with probability $1-p$).  Here, $L$ 
is the number of spins in the spin chain, which is equal to the number of steps on the dual lattice required to traverse the network from left to right.  The time $t$ is equal to the number of layers of unitary operations applied, which is the number of steps required to traverse the network from top to bottom.  (For example, Fig.\ \ref{fig:perclatticemapping} shows $L = 6$ and $t = 5$.)  For a given random realization of the network, and for a given choice of subsystem, we search deterministically for the minimal-cost pathway that separates the subsystem from the rest of the network.  (Such a pathway may, in general, extend outside the network, and may not be unique.)  $S_0$ is defined as the total cost of this minimal-cost pathway, which corresponds to the number of unbroken bonds that must be cut in the circuit in order to separate the subsystem from the rest of the network.  All data presented in this section are averaged over many random realizations of the network for each choice of parameters.

In Fig.\ \ref{fig:validation} we check the results of this approach against data from a full matrix product state simulation of the random unitary circuit (we modify the boundary condition of the percolation problem to match those used in the quantum simulation, Sec.~\ref{sec:models}). One can see that the two sets of data are very close.  A small systematic error appears in the quantum estimation of $S_0$ due to spurious small eigenvalues of the reduced density matrix that arise from numerical truncation error.  This issue is discussed further in Appendix \ref{sec:quantumsimulation}.

In the remainder of this section, all results are taken from the classical percolation network simulation, which is significantly faster computationally.

\subsection{Universal dynamics of $S_0(t)$}
\label{sec:universalS0dynamics}

\begin{figure}[t]
\begin{center}
\includegraphics[width=0.77\linewidth]{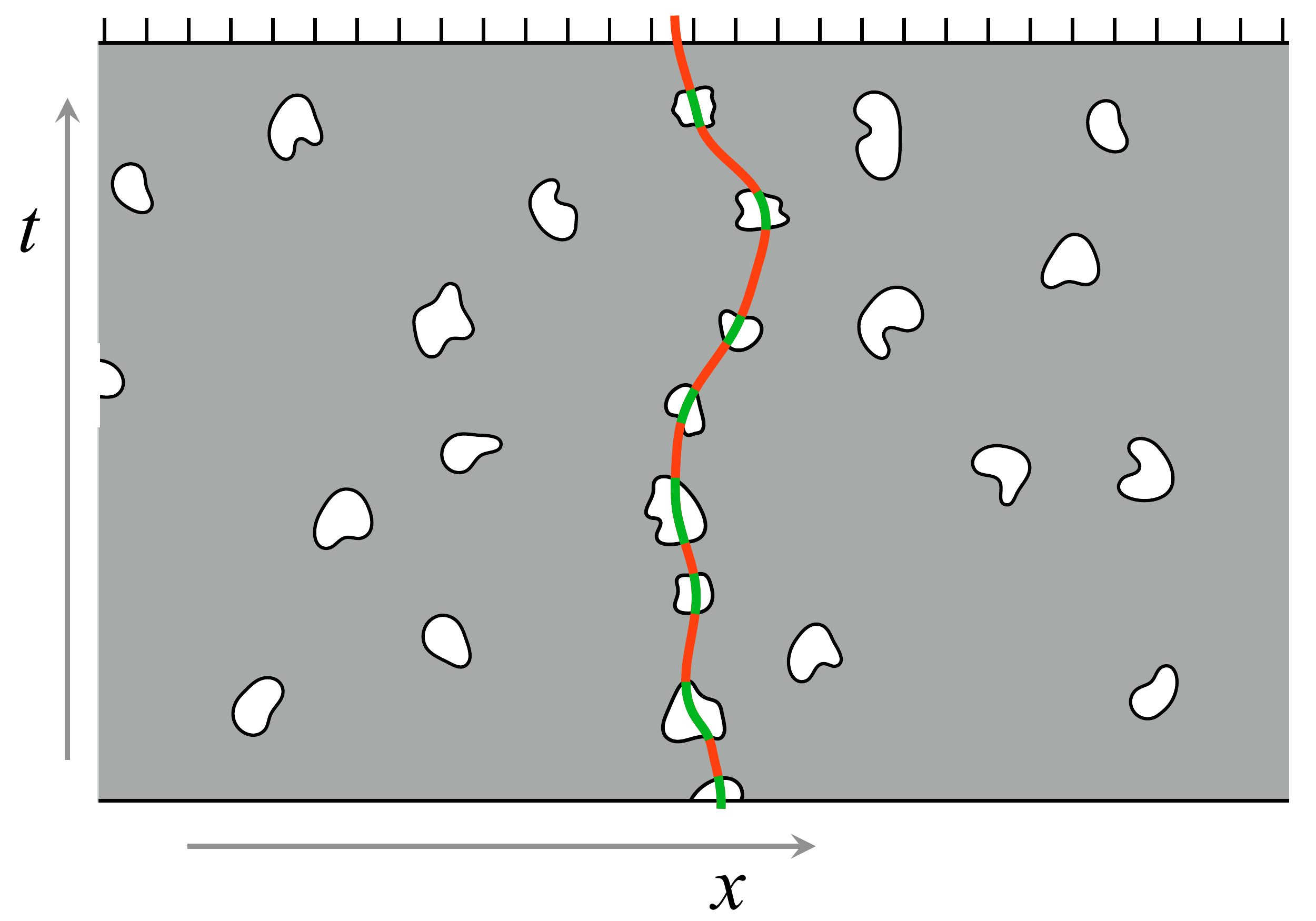}
\end{center}
\caption{Cartoon of the minimal cut in the \textit{entangling} phase for $S_0$
(at $L=\infty$).
The small white domains are clusters of broken bonds: we only show those clusters which the minimal cut passes through.
A minimal cut is shown traversing the spacetime patch. The red sections have nonzero cost per unit length while the green regions have zero cost.
In the entangling phase the cost of the minimal cut scales like $t$ (with subleading KPZ fluctuations), so that $S_0(t)\sim t$.}
\label{fig:percentanglingphase}
\end{figure}

Given the percolation picture, one can understand the dynamics of  the Hartley entropy $S_0$ as a function of time $t$ using the following scaling arguments.
Let the initial state at time zero be unentangled.
We first study the growth of entanglement, between two halves of the system, in the limit where the  system size $L\rightarrow \infty$.
In this case the minimal cut meanders from the top to the bottom of the circuit, as illustrated in Fig.\ \ref{fig:perclatticemapping}.\footnote{Note that, when $p$ is nonzero, the minimal cut is not guaranteed to be directed. By contrast, in the case of a purely unitary circuit without measurements, the minimal cut (which may not be unique) can  always be chosen to be directed. This is a consequence of causality in the unitary circuit, which means that only the unitaries within a lightcone can affect the entanglement. This lightcone structure is lost in the presence of measurements.}

One can understand the existence of two phases for $S_0$ by considering the properties of this cut deep in either of the two phases, where $p$ is either close to zero or close to unity.

At small $p$, broken bonds exist only in small, isolated clusters. The minimal cut is therefore forced to pass through a number of unbroken bonds that is proportional to $t$ with an $\mathcal{O}(1)$ coefficient. This situation is shown in Fig.~\ref{fig:percentanglingphase}.
In this phase randomness in the location of measurements has a subleading effect on the entanglement,  giving subleading Kardar-Parisi-Zhang fluctuations \cite{kardar_dynamic_1986,HuseHenleyFisherRespond}, as for purely unitary dynamics with randomness \cite{nahum_quantum_2017, zhou2018emergent, knap2018entanglement}.

\begin{figure}[t]
\begin{center}
\includegraphics[width=0.8\linewidth]{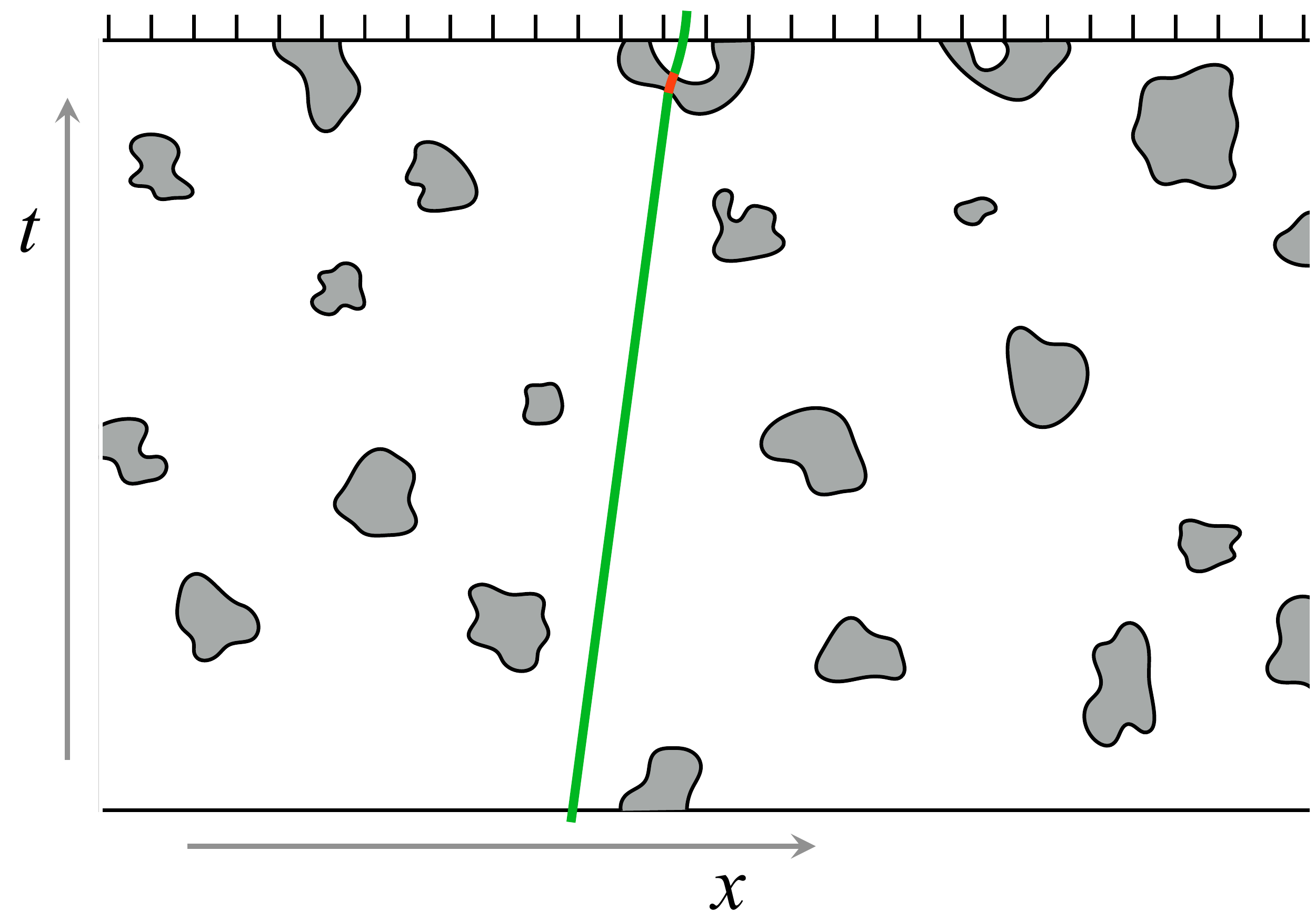}
\end{center}
\caption{Cartoon of the minimal cut in the \textit{disentangling} phase for $S_0$  (cf. Fig.~\ref{fig:percentanglingphase}).
The clusters of broken bonds now percolate, forming the infinite white region. The cost of the minimal cut remains finite as $t\rightarrow \infty$: only a finite number of unbroken bonds need to be traversed to reach the infinite white cluster. Therefore $S_0(t)\sim t^0$.}
\label{fig:percdisentanglingphase}
\end{figure}

At $p$ close to unity, on the other hand, broken bonds predominate over unbroken bonds, and there is a  connected cluster of unbroken bonds that spans the entire system.
Thus at $p > p_c$ the only contribution to the entanglement comes from passing through unbroken bonds located near the starting point of the cut.  (For the bipartite entanglement, this starting point is at the top center of the network.)  Once this small set of unbroken bonds has been traversed, the minimal cut can proceed arbitrarily far in the time direction using only broken bonds, and consequently the average $S_0$ becomes independent of $t$ at large $t$.  This is illustrated in Fig.\ \ref{fig:percdisentanglingphase}.

Let us now consider scaling at and near the critical point.  The percolation threshold for the square lattice is $p_c = 1/2$.
At $p = p_c$ the clusters of broken bonds have a scale-invariant fractal structure.
A key point is that adjacent clusters of some large size $\sim \ell$ approach each other to a distance of one lattice spacing (in fact clusters which approach each other to unit separation do so at $\sim \ell^{3/4}$ different places \cite{coniglio1989fractal}). This means that the minimal cut can pass from one large ``empty chamber'' to an adjacent one, of similar scale, for unit cost.

\begin{figure}[t]
\begin{center}
\includegraphics[width=0.75\linewidth]{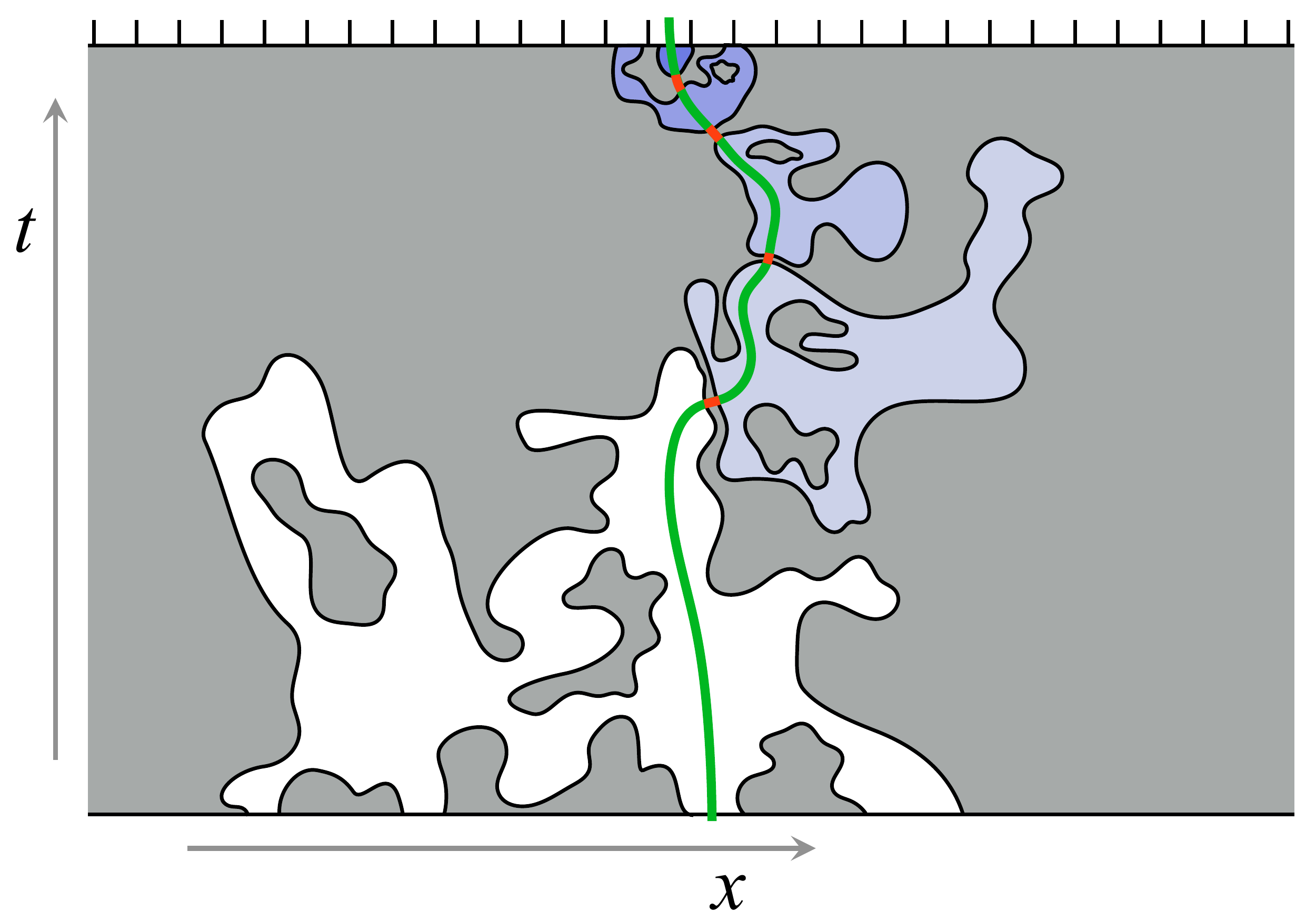}
\end{center}
\caption{Cartoon for the scaling argument showing ${S_0(t) \sim \log t}$ at the percolation critical point (cf. Figs.~\ref{fig:percentanglingphase},~\ref{fig:percdisentanglingphase}). The minimal cut passes through the sequence of white domains shown in blue/white. Writing the linear sizes of consecutive domains in this sequence as $R_1, R_2, \ldots$, the ratio $R_{i+1}/R_i$ is typically larger than one  (see text), so for an $i$ of order $\log t$ the minimal cut reaches a cluster of $\mathcal{O}(t)$ size  that borders the boundary and the sequence ends. Domains $i$ and $i+1$ typically approach each other to within one lattice spacing, so the cost scales as the number of domains in the sequence.}
\label{fig:perccriticalscaling}
\end{figure}

At the top of the network, where the cut is anchored, the minimal path typically passes into an empty chamber of size $\mathcal{O}(1)$. It then passes though a sequence of larger chambers until it reaches one of scale $\sim t$ and exits the system. Typically, at each step, the path can find a chamber that is larger by an $\mathcal{O}(1)$ factor than the previous one.
This progression is illustrated schematically in Fig.\ \ref{fig:perccriticalscaling}.
 Therefore the total number of chambers in the sequence is only of order $\ln t$, and
\be
\label{eq:S0critp}
S_0(t, p_c) \simeq  A\, \ln t.
\ee
This logarithmic scaling for critical first passage is proved in Ref.\ \onlinecite{chayes1986critical}.
The coefficient $A$, which can be thought of as an entanglement per scale, is universal as a result of the scale-invariance of the process. Below, we estimate $A=0.27(1)$.

The typical size $\xi$ of the empty chambers is finite for $p\lesssim p_c$, but diverges with the correlation length exponent as $p_c$ is approached:
\ba\label{eq:correlationlength}
\xi &\sim 1/|p - p_c|^{\nu}.
\end{align}
For classical percolation in two dimensions, the critical exponent $\nu = 4/3$.
The part of the minimal cut within $\xi$ of the top of the sample costs $A \ln \xi$, as above. At farther distances from the starting point the path travels through chambers of size $\sim \xi$. Each chamber-to-chamber crossing involves passing through an order-unity number of unbroken bonds, and consequently the minimal cut passes through a total number $S_0 \sim t/\xi$  of unbroken bonds, so that:
\ba
S_0(t,p) & \sim (p_c - p)^{\nu} \times t, &  (p & \lesssim p_c).
\label{eq:S0smallp}
\end{align}
That is, the entanglement $S_0$ at $p < p_c$ grows without bound as a function of time, with a growth rate that vanishes as $p \rightarrow p_c$.

At $p \gtrsim p_c$ the minimal cut escapes to the infinite cluster of unbroken bonds after a section of length $\mathcal{O}(\xi)$ at the top of the lattice. On scales smaller than $\xi$ the critical scaling applies, giving a cost $A \ln\xi$, so that
\ba
\label{eq:S0largep}
S_0(t, p) & \simeq A \, \nu \, \ln\left( \frac{1}{p - p_c} \right) &  (p & \gtrsim p_c).
\end{align}
The above results are all limiting cases of the general scaling form
\be
\label{eq:scalingformt}
S_0(t, p) = A \ln \xi + F(t/\xi),
\ee
which can be obtained by imagining rescaling both $t$ and $\xi$ by a constant factor and repeating the considerations above. The asymptotics of $F$ for large and small argument are determined by Eqs.~(\ref{eq:S0smallp}) and (\ref{eq:S0largep}).
Generalizations of Eq.~(\ref{eq:scalingformt}) will be useful to us below in extracting exponents numerically.
This sort of scaling form (with length in place of time) has also been derived for scaling of higher R\'{e}nyi entropies in critical random tensor network states \cite{vasseur2018entanglement}, by a different kind of reasoning.

The scaling results of Eqs.\ (\ref{eq:S0critp})--(\ref{eq:scalingformt}) can be easily generalized to the case where the system size is finite and the time $t \gg L$.  In this case, the minimal cut travels a total horizontal distance $L/2$, moving from the top-center of the lattice to one of the lateral edges.
The above results hold if we replace $t$ with $L$ in Eqs.\ (\ref{eq:S0critp})--(\ref{eq:scalingformt}), except that the scaling function $F$ in Eq.~(\ref{eq:scalingformt}) is different. If $t$ and $L$ are both finite, the scaling form has an additional dependence on $t/L$. In general a nonuniversal speed $v$ would enter, but here that is fixed to unity by the symmetry of the square lattice between time-like and space-like directions. We comment further on the symmetry between $t$ and $L$ in the following section.

\begin{figure}[htb]
\begin{center}
\includegraphics[width=\linewidth]{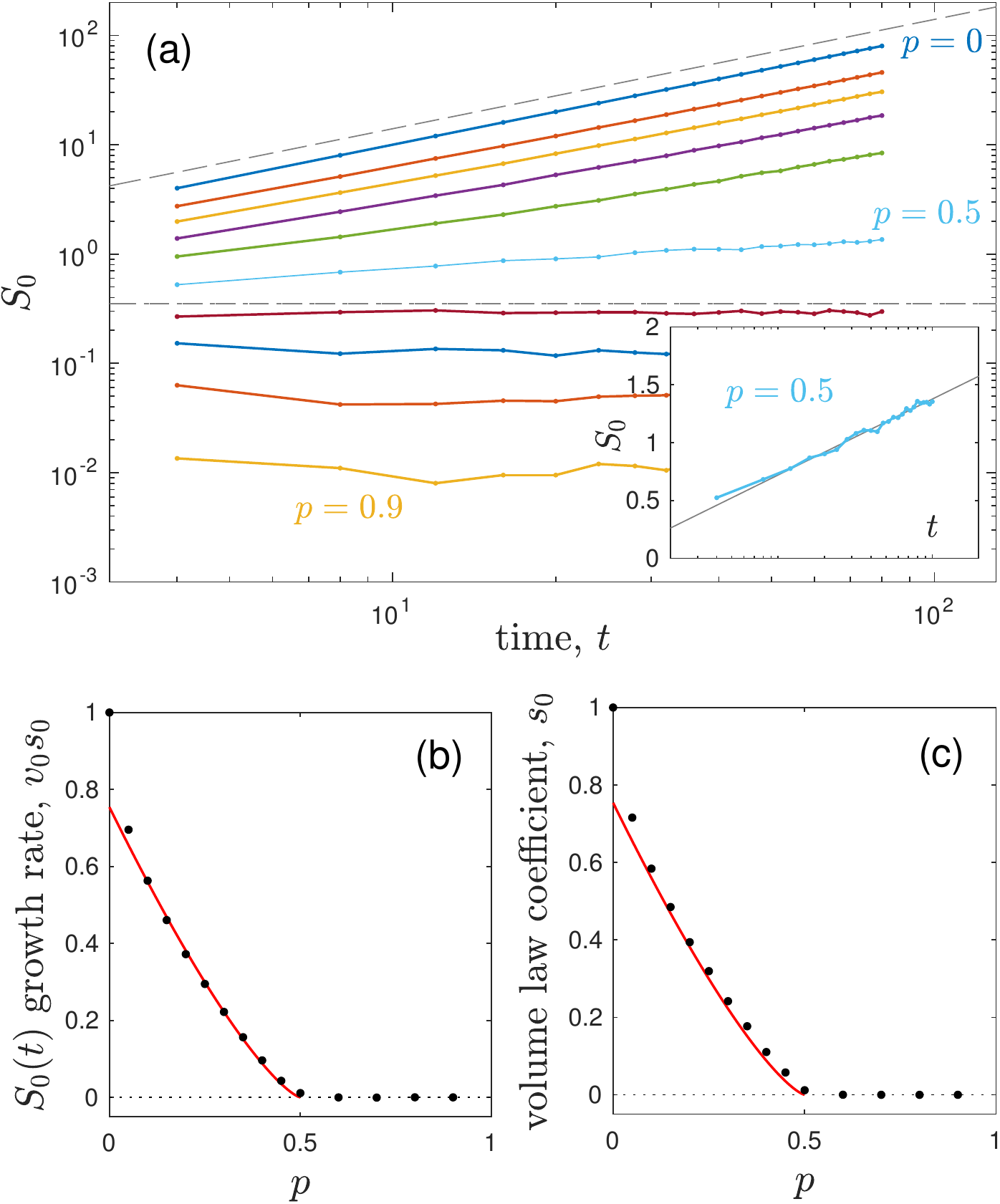}
\end{center}
\caption{Growth of the minimal cut cost $S_0$ separating two halves of the classical percolation network, as measured by numerical simulations.  (a) shows the dependence of $S_0$ on the time $t$ for a network with aspect ratio $L = 4t$. (Note the double logarithmic scale).  Different curves correspond to different values of $p$, ranging from $p = 0$ (topmost curve) to $p = 0.9$ (bottom curve) in steps of $0.1$.  The dashed lines show, respectively, the dependence $S_0 \propto t$ and $S_0 = \textrm{const.}$  The inset shows $S_0$ at $p = 0.5$ on a semi-logarithmic scale, and the thin solid line shows a fit to the form $S_0 = A \ln t + B$, which gives $A = 0.27 \pm 0.01$ and $B = 0.063$. (b) shows the rate of entanglement growth $v_0 s_0$ as a function of $p$, extracted by measuring the linear slope of the data in (a).  The red line shows $v_0 s_0 \propto (p-p_c)^{4/3}$, as suggested by Eq.\ (\ref{eq:S0smallp}).  (c) shows the coefficient of the volume law entanglement, $s_0$, as a function of $p$, which is extracted by measuring the linear slope of the dependence $S_0(L) \propto s_0 L$ for networks with $t = 4L$. The red line is identical to the one in (b).  All data in this figure is averaged over 4000 realizations of the random network.
}
\label{fig:S_growth}
\end{figure}

The scaling expectations above can be tested using our numerical simulations.
First, Fig.~\ref{fig:S_growth}(a) shows $S_0(t)$  for a range of values of $p$ (in a system of fixed aspect ratio, ${L=4t}$).  As expected, $S_0$ grows linearly in $t$ when ${p < 1/2}$, and remains constant in the limit of large $t$ at ${p > 1/2}$.  The inset shows that our data is also consistent with the logarithmic dependence $S_0 \propto \ln t$ at ${p = 1/2}$.

In the entangling phase we may define an asymptotic entanglement growth rate (for a quench from an area law state) as well as an entropy density $s_0$ associated with the volume law entanglement after saturation. The latter is given by
\be
{S_0(t \rightarrow \infty, L) \simeq s_0 \times \frac{L}{2}}.
\ee
The growth rate is given by
\be
{S_0(t, L\rightarrow \infty) \simeq v_0 s_0 \times t},
\ee
where (by definition)  $v_0$ is the `entanglement speed' for the zeroth R\'{e}nyi entropy.  The quantities $v_0 s_0$ and $s_0$ are plotted as functions of $p$ in Figs.\ \ref{fig:S_growth}(b) and (c).  Both vanish as $p\rightarrow p_c$, consistent with Eq.\ (\ref{eq:S0smallp}) (and the analogous equation for the opposite regime of aspect ratio)
from which we expect 
\be
s_0\sim (p_c-p)^{4/3},
\ee
for the entropy density at $p\lesssim p_c$.  The equivalence between $v_0 s_0$ and $s_0$ as a function of $p$ [seen in Figs.\ \ref{fig:S_growth}(b) and (c)], suggests an entanglement velocity $v_0 \sim 1$, which is implied by the symmetry of the lattice.

More generally, in the entangling phase one can understand the coarse-grained dynamics of the entanglement using a coarse-grained minimal membrane picture similar to that in the unitary case \cite{nahum_quantum_2017, jonay2018coarse, mezei_membrane_2018}.  This picture applies to more general initial states or more general choices of subsystem, and in general dimensionality.

\begin{figure}[t]
\begin{center}
\includegraphics[width=\linewidth]{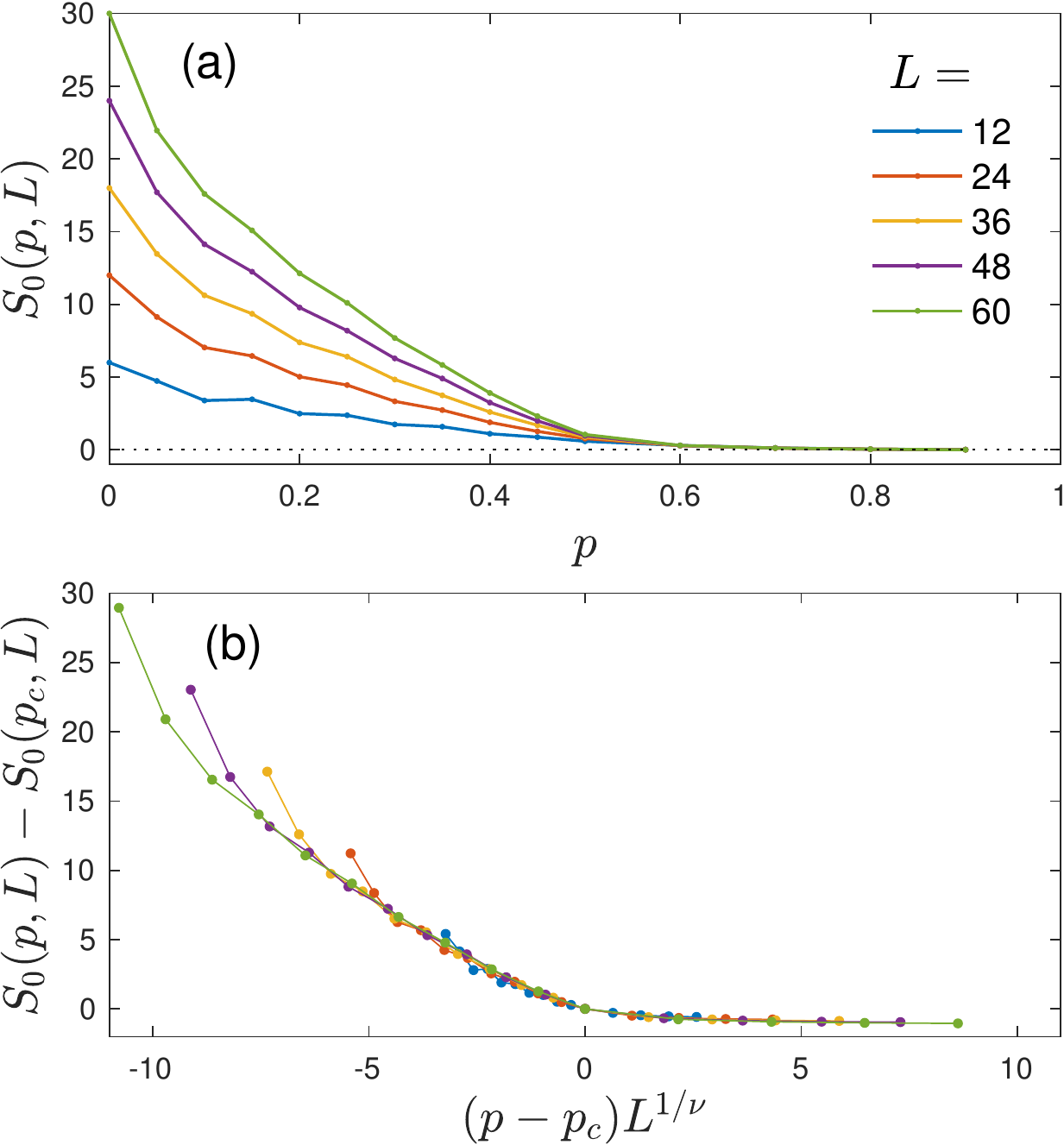}
\end{center}
\caption{Demonstration of critical behavior in the zeroth R\'{e}nyi entropy near the percolation threshold $p_c = 1/2$.  (a) shows $S_0$ as a function of $p$ for different values of the system size $L$, with $t = 4L$.  (b) shows critical scaling of this same data, with $\nu = 4/3$ and $p_c = 1/2$ taken from the known values for two-dimensional percolation on the square lattice.  The vertical axis is $S_0(p,L) - S_0(p_c,L)$, where $S_0(p_c,L)$ is the measured value of $S_0$ at $p = 1/2$ for a given value of $L$.
}
\label{fig:S0_scaling}
\end{figure}

Let us now consider how to make use of the scaling form Eq.~(\ref{eq:scalingformt}) to extract $p_c$ and $\nu$ in numerics.
In a system of fixed aspect ratio  (we now take $t=4L$) we have the scaling form $S_0(t, p) = A \ln L + G(L/\xi)$, as noted above. This means that if we subtract $S_0(t,p_c)$ from $S_0(t,p)$ we obtain a  pure scaling function, without the logarithmic term:
\be
S_0(L,p) - S_0(L,p_c) = \widetilde G \big( (p-p_c)L^{1/\nu} \big).
\ee
 We should therefore see a scaling collapse if we plot the left hand side as a function of $(p-p_c)L^{1/\nu}$.
Such a collapse is demonstrated in Fig.\ \ref{fig:S0_scaling}.

In the present case $p_c$ and $\nu$ are known analytically, but this kind of scaling collapse will be useful in the next section, where the analogous quantities must be determined empirically.
To get an idea of the precision of this process, here one can also attempt to do an unbiased search for the values of $p_c$ and $\nu$ that produce the best scaling collapse of the data.  Our algorithm for this search is described in Appendix \ref{sec:scalingdetails}.  Briefly, this algorithm seeks to minimize an objective function that is equal to the sum of the square residuals of all curves $S_0(p,L) - S_0(p_c,L)$ from their common mean at a given value of $(p-p_c)L^{1/\nu}$, summed over all unique values of $(p - p_c)L^{1/\nu}$ that are present in the data set. A simple gradient descent search returns the values of $p_c$ and $\nu$ that minimize this objective function.

Performing such a search using the data in Fig.\ \ref{fig:S0_scaling}(a) yields $p_c = 0.51 \pm 0.01$ and $\nu = 1.24 \pm 0.13$, as listed in Table \ref{tab:exps}.
Closely matching results are obtained if one analyzes data for $S_0$ obtained from the `quantum' simulations of the random unitary circuit or Floquet dynamics, which are restricted to smaller size. These scaling collapses are shown below in  Fig.\ \ref{fig:all_scaling}(a)--(b), and Table \ref{tab:exps}, and all are consistent with the exact values $p_c = 1/2$ and $\nu = 4/3$ for two-dimensional percolation.
In Sec.~\ref{sec:genericdynamicaltransition} we apply our algorithm for data collapse to the higher-order R\'{e}nyi entropies $S_1$, $S_2$ and $S_\infty$ near the generic transition, where no such exact values are available.

\subsection{Spatial correlations \& long-range entanglement}
\label{sec:I0correlations}

\begin{figure}[t]
\begin{center}
\includegraphics[width=0.8\linewidth]{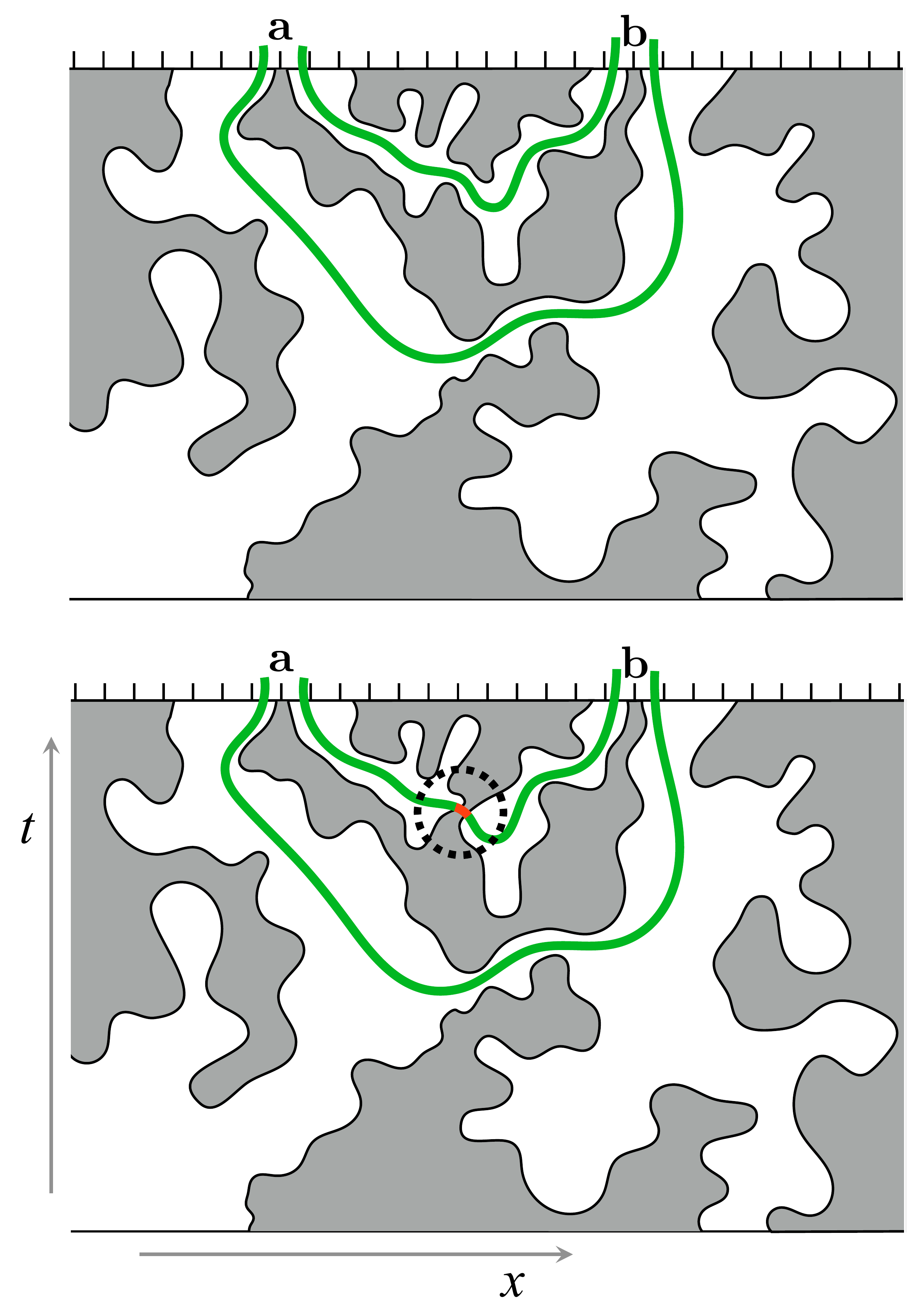}
\end{center}
\caption{Examples of minimal cut configurations for $S_0({\bf a}\cup {\bf b})$ in cases where $I_0=2$ (Upper) and $I_0 = 1$ (Lower).
In the former case the cost of the minimal cut $S_0({\bf a}\cup {\bf b})=0$ and in the latter case it is $S_0({\bf a}\cup {\bf b})=1$, i.e the obstacle marked in the Upper figure is of minimal width.
In both cases  $S_0({\bf a})=S_0({\bf b})=1$.}
\label{fig:I0mincut}
\end{figure}

The critical stationary state at $p_c$ has power-law spatial correlations that reflect its unusual scale-invariant entanglement structure.
The simplest measures of these correlations are the R\'{e}nyi mutual informations, $I_n$, between a pair of distant spins  ${\bf{a}}$ and ${\bf{b}}$ (Sec.~\ref{sec:models}):
\be\label{eq:I0defn2}
I_n({\bf{a}}, {\bf{b}}) = S_n({\bf{a}}) + S_n({\bf{b}}) - S_n({\bf{a}} \cup {\bf{b}}).
\ee
In this section we discuss $I_0$, which is critical at the percolation critical point. The quantities $I_n$ with higher $n$ are instead critical at the generic entanglement transition that occurs at smaller $p$.
$I_0$ can be computed numerically using minimal cut configurations with three different boundary conditions corresponding to the terms in Eq.~(\ref{eq:I0defn2}).

Let $I_0(x)$ be the mutual information for a pair of spins separated by a distance $x$, averaged over realizations. For simplicity consider an infinite spin chain that has been evolved for infinite time, and so is in the steady state corresponding to a particular $p$.

Note first that $I_0(x)$ decays exponentially with $x$ on \textit{both} sides of the critical point.
The disentangled state is `close' to a product state, and connected correlations fall off rapidly with distance.
In the entangling phase subsystems are strongly entangled with their exterior, but the mutual information shared between any \textit{two} small subsystems is negligible: the average of $S_0({\bf{a}} \cup {\bf{b}})$ is exponentially close to that of $S_0({\bf{a}}) + S_0({\bf{b}})$.
For example, consider $p=0$: the reduced density matrices thermalize to `infinite temperature', and all local correlations vanish.
However at the critical point neither of these mechanisms destroy correlations.

\begin{figure}[t]
\begin{center}
\includegraphics[width=\linewidth]{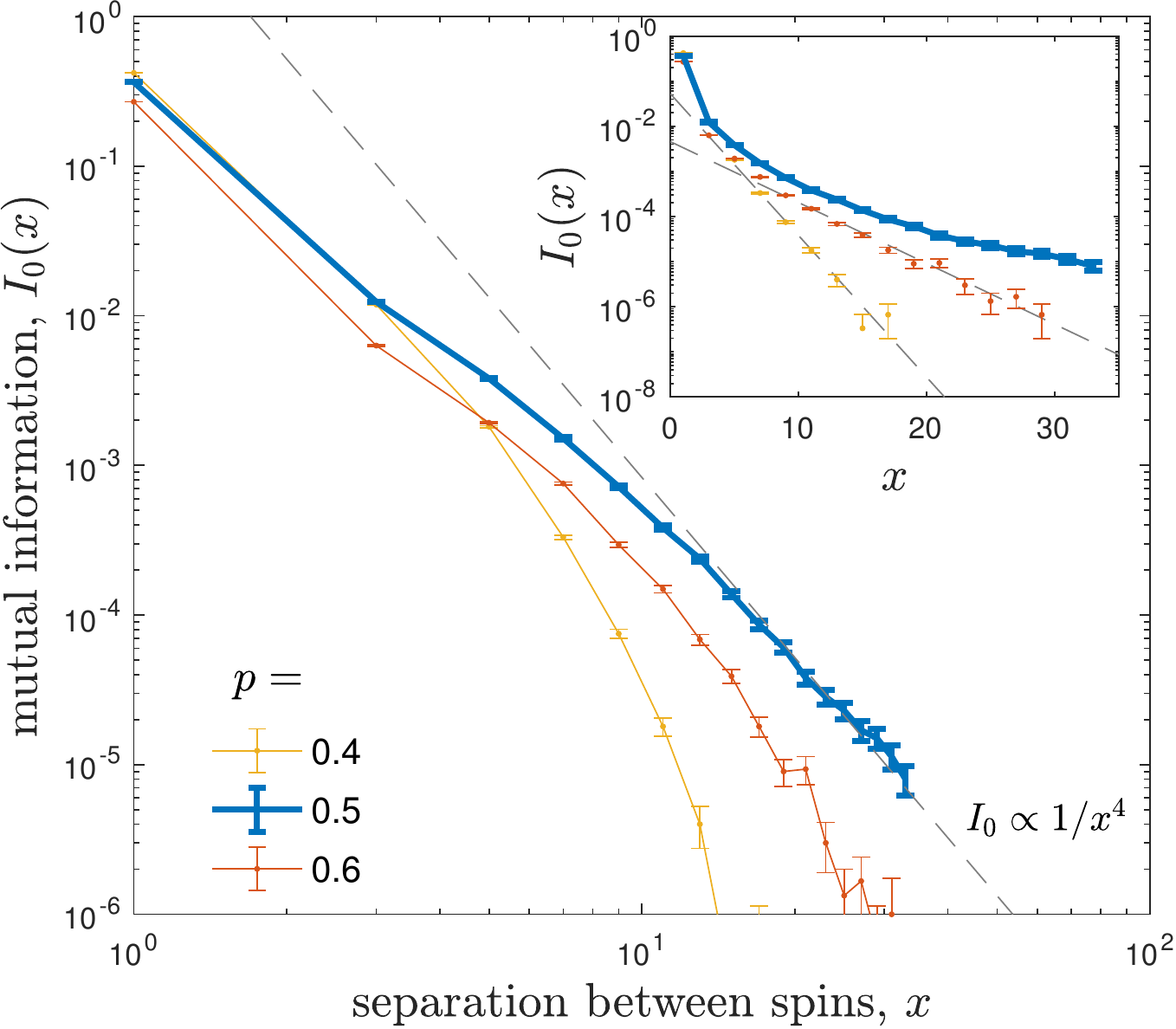}
\end{center}
\caption{Decay of the mutual information $I_0$ between two spins as a function of their separation $x$, as determined by the classical percolation picture.  The main panel shows $I_0(x)$ for three different values of $p$.  The dashed line indicates the dependence $I_0 \propto 1/x^4$ expected for $p = p_c$ at large $x$.  The inset shows the same data plotted on a semilogarithmic scale.  The dashed straight lines in the inset indicate exponential decay, which describes the curves at $p \neq p_c$.  All data corresponds to system size $L = 3x$, with the two spins located at positions $L/3$ and $2L/3$.  Data is averaged over $3\times 10^6$ random realizations and the evolution time is $t = L$. Error bars indicate one standard deviation divided by the square root of the number of realizations.}
\label{fig:I0_figure}
\end{figure}

The mutual information for a pair of spins separated by distance $x$, $I_0(x)$, is plotted in Fig.~\ref{fig:I0_figure}. (We use a chain of length $L=3x$ that has been evolved for a time $t=L$, rather than an infinite system, but this does not change the exponent below.) As expected there is exponential decay for $p>p_c$ and $p<p_c$. But at $p=p_c$ the data at large distances is consistent with
\ba
\label{eq:percolationcorrelator}
I_0(x) & \propto \frac{1}{x^{2\Delta_c}},
&
\Delta_c & = 2.
\end{align}
This exponent is exact and follows from known results for percolation \cite{duplantier1987exact,rushkin2007critical,dubail2010conformal}, as discussed next.
Our percolation mapping also gives an intuitive picture for the scale-invariant entanglement structure underlying Eq.~(\ref{eq:percolationcorrelator}).

In more detail, there are three possible values for the mutual information in a given realization: ${I_0(x)=0,1, 2}$. At large $x$, almost all configurations give $I_0(x)=0$.
The probability of either of the nonzero values is $\mathcal{O}(x^{-4})$.
Situations that give finite $I_0$ are like those shown in Fig.~\ref{fig:I0mincut}.
It is simplest to consider cases where $I_0=2$.
These occur  when the topology of the percolation configuration is as shown in Fig.~\ref{fig:I0mincut} (Upper): there is a percolating cluster of unbroken bonds connecting the two spins, and this cluster does not touch the boundary elsewhere.
The probability of such a configuration scales as above \cite{duplantier1987exact,rushkin2007critical,dubail2010conformal}.\footnote{$\Delta_c=2$ is the scaling dimension of the appropriate boundary operator in percolation.}
A configuration with $I_0=1$ can be obtained by adding a narrow bridge of occupied bonds as in  Fig.~\ref{fig:I0mincut} (Lower): one can argue%
\footnote{This follows from the fact that the typical number of configurations $N_{1\rightarrow 2}$ of $I_0=2$ type that can be reached from a given configuration of $I_0=1$ type, and the number $N_{2\rightarrow 1}$  of configurations of $I_0=1$ type that can be reached from a given configuration of $I_0=2$ type, scale with the same positive power of $x$ (given by the fractal dimension of the `red bonds' \cite{coniglio1989fractal}).}
 that the probability of such a configuration also scales as $x^{-2\Delta}$. This is borne out by our numerics: the ratio of the probabilities of the two values of $I$ tends to an $\mathcal{O}(1)$ constant as $x\rightarrow\infty$.

\section{The generic dynamical transition}
\label{sec:genericdynamicaltransition}

For the entropies $S_n$ with $n > 0$, there is no mapping to classical percolation, at least in a generic model.  
Nonetheless, we find that many qualitative features of the toy model carry over to the physical entanglement dynamics, including the existence of a finite threshold $p_c$ separating entangling and disentangling phases,
and a nontrivial scale-invariant state at $p_c$. 
We show below that the threshold $p_c$ is lower than that in the toy model: for example in the random unitary circuit we find ${p_c = 0.26 \pm 0.08}$, compared to $p_c=1/2$ in the toy model.
Therefore there is a regime of measurement rate given by ${p_c < p < 1/2}$  in which $S_0$ grows linearly with time and the ``minimal cut" picture suggests a volume-law entanglement, 
but the von Neumann and higher-order entropies in fact obey area-law scaling.\footnote{%
The existence of a regime $p_c < p < 1/2$ has the following implication about the distribution of eigenvalues $\{ \lambda_i \}$ of the reduced density matrix $\rho_A$. Equation (\ref{eq:Sndef}) can be rewritten in terms of these eigenvalues as $S_n = [1/(1-n)] \log_2 \left( \sum_i \lambda_i^n \right)$.  In the regime $p_c < p < 1/2$, the volume-law behavior of $S_0$ implies that the number of nonzero eigenvalues diverges in the limit of infinite system size, while the area-law behavior of $S_n$ with $n \geq 1$ implies that $\sum_i \lambda_i^n$ remains finite for all $n \geq 1$.
}
We also find an exponent $\nu$ that is different from $4/3$, implying that the critical behavior for  $S_n$ with $n \geq 1$ is in a universality class that is distinct from two-dimensional percolation.

\subsection{Dynamics of $S_n$ at the generic transition} 

Our main tool for studying the dynamics of the entanglement is a numerical simulation of the unitary circuit with measurements (we use the ITensor package to manipulate matrix product states \cite{ITensor}).  Details of these simulations are provided in Ref.\ \onlinecite{nahum_quantum_2017} for the case without measurements. 
We modify these simulations only by the stochastic application of projective measurements after each unitary operation. 
The outcome of each measurement is chosen randomly with a probability determined by the Born rule.  For each value of $p$ and each type of simulation dynamics (either random unitary or Floquet), results for the entanglement are averaged over many random realizations.

\begin{figure}[t]
\begin{center}
\includegraphics[width=\linewidth]{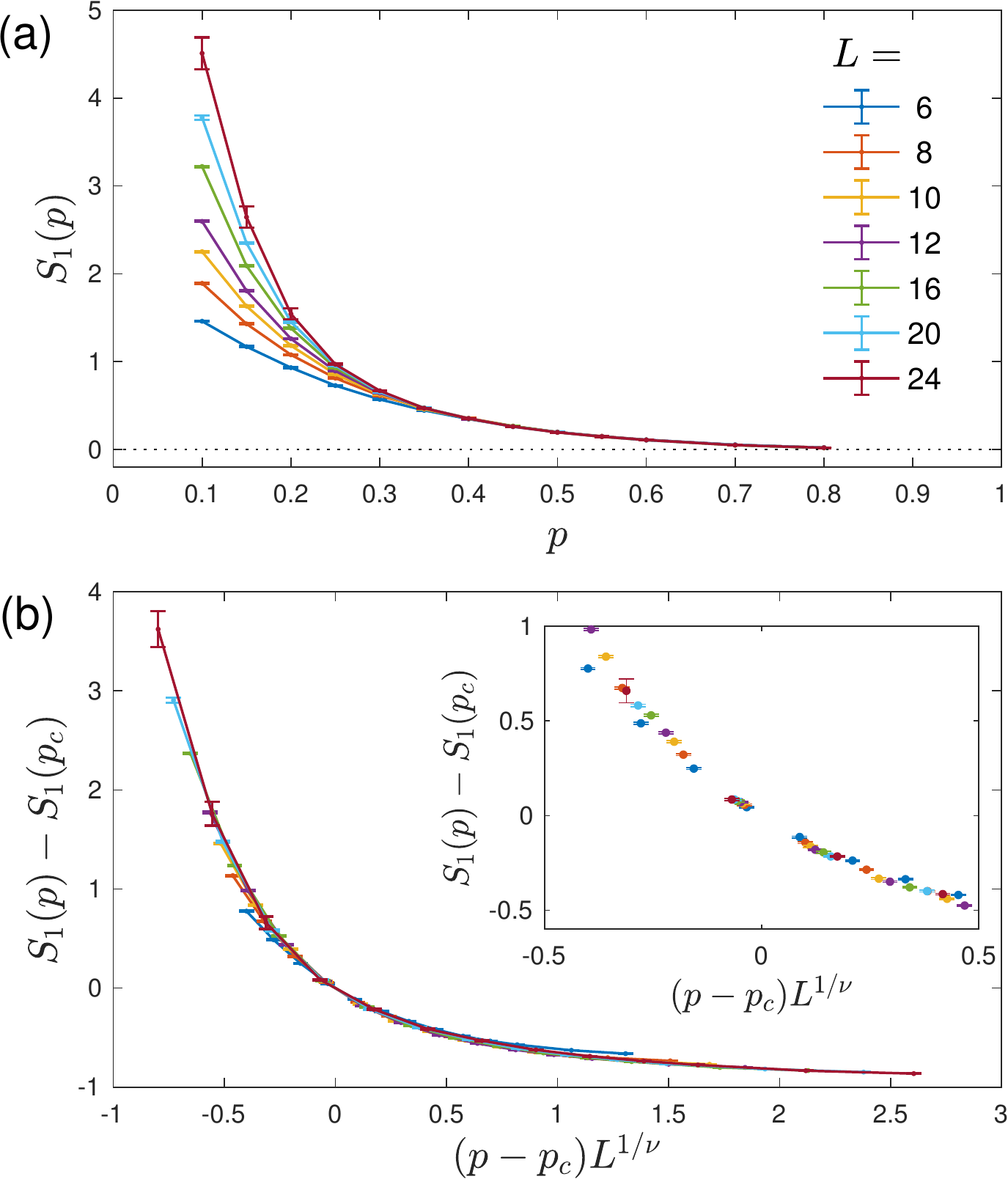}
\end{center}
\caption{Critical behavior of the von Neumann entanglement entropy $S_1$ for the random unitary circuit.  (a) shows $S_1$ as a function of $p$ for different values of the system size $L$, with $t = 2 L$.  (b) shows critical scaling of this same data.  The inset shows a zoomed-in view of the same data near the origin (with the solid lines suppressed).
}
\label{fig:S1_scaling}
\end{figure}

An example of our simulation results is plotted in Fig.\ \ref{fig:S1_scaling}(a), which shows the von Neumann entropy $S_1$ for the case of random unitary dynamics, as a function of $p$, for different system sizes. At small $p$, the value of $S_1$ depends strongly on system size, suggesting a volume-law entanglement.  At large $p$, however, data from different system sizes collapse onto a single value, suggesting area-law behavior.

We demonstrate that there is indeed a transition between two phases using a scaling collapse  of the type described at the very end of Sec.~\ref{sec:universalS0dynamics}.
We showed in Sec.\ \ref{sec:zerothRenyi} that at the transition of the zeroth Renyi entropy we have the scaling form $S_0 = A \ln t + F(u t/\xi, u t/L)$  (here $u$ is a nonuniversal speed, which in the lattice model of Sec.\ \ref{sec:zerothRenyi} was fixed to $u=1$ by symmetry). Our results for the generic transition are consistent with the same scaling form for the higher entropies, $S_n = A_n \ln t +F_n(v t/\xi, u t/L)$. The value of $p_c$ is different for the generic transition, as are universal constants such as the correlation length exponent $\nu$; we also allow for dependence of $A_n$ and $F_n$ on the R\'enyi index $n$.
We give evidence below that the dynamical critical exponent at the transition, $z$, is unity, so that $x$ and $t$ behave the same way under rescaling and our assumption that $t/L$ is the appropriate scaling variable is justified.

This scaling form implies that for a system of fixed aspect ratio, the difference
\be
S_n(L,p) - S_n(L,p_c)  \nonumber
\ee
depends on $L$ and $p$ only through the combination $(p-p_c)L^{1/\nu}$, which enables us to perform a numerical scaling collapse. 
The values of $p_c$ and $\nu$ are estimated using the algorithm described in Sec.\ \ref{sec:universalS0dynamics} and detailed in Appendix \ref{sec:scalingdetails}. 

The resulting scaling collapse is shown in Fig.\ \ref{fig:S1_scaling}(b) for the von Neumann entropy in the case of random unitary dynamics.  As noted in Table \ref{tab:exps}, the values of $p_c$ and $\nu$ obtained are $p_c = 0.26 \pm 0.08$ and $\nu = 2.01 \pm 0.10$.

\begin{figure}[t]
\begin{center}
\includegraphics[width=\linewidth]{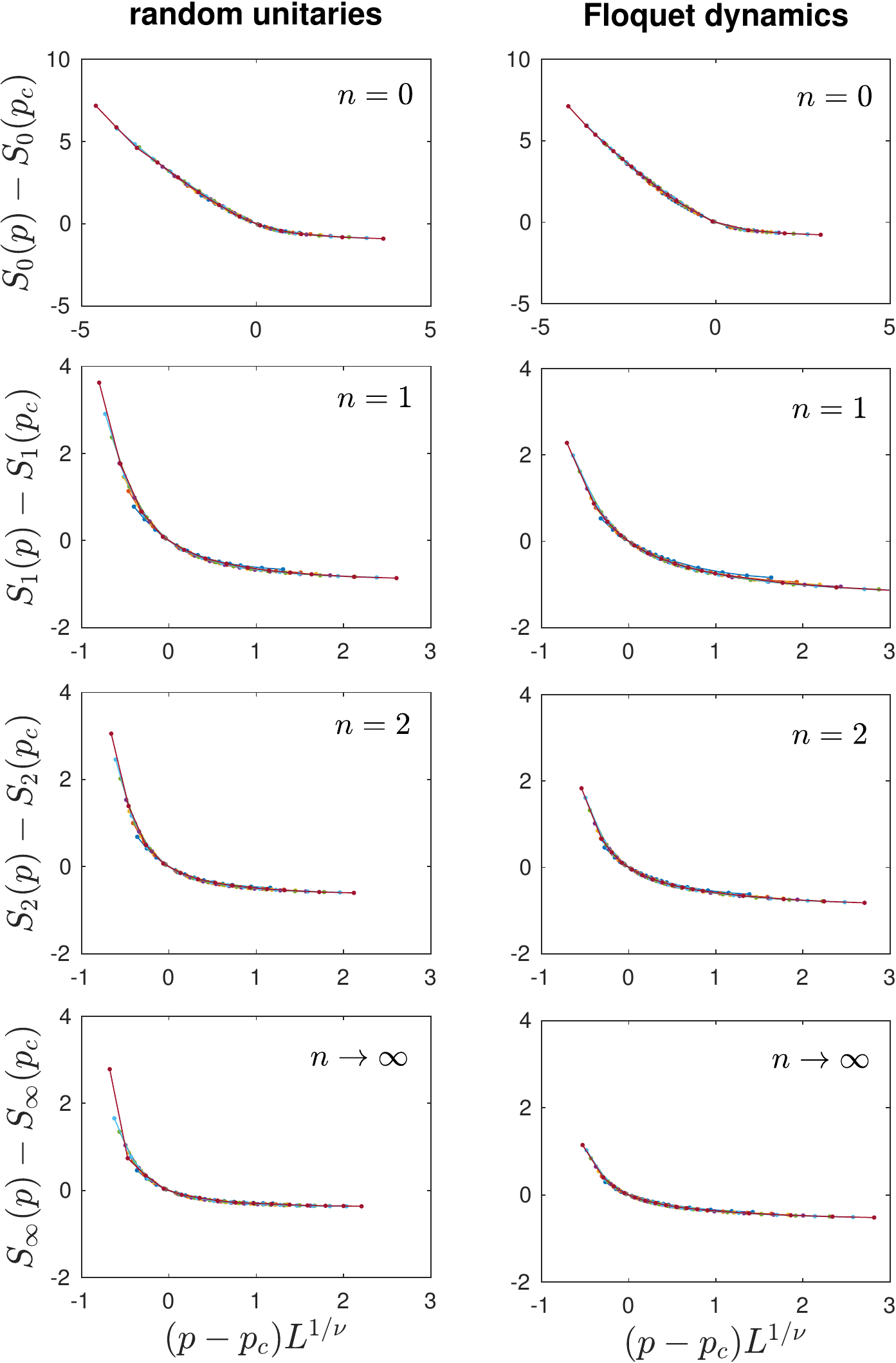}
\end{center}
\caption{Critical scaling for the bipartite entanglement $S_n$ as measured by numerical simulations of the random unitary circuit (left column of plots) and the Floquet dynamics (right column).  Each plot is labeled by the corresponding value of $n$, and shows curves for seven different system sizes, ranging from $L = 6$ to $L = 24$ [the same as in the legend of Fig.\ \ref{fig:S1_scaling}(a)].  The simulation time was $t = 2L$ for the simulations of random unitaries, and $t = 8L$ for the Floquet dynamics.  The corresponding values of $p_c$ and $\nu$ for each plot are listed in Table \ref{tab:exps}.
}
\label{fig:all_scaling}
\end{figure}

The full set of scaling collapses, for both types of dynamics and for R\'{e}nyi indices $n=1, 2, \infty$ (and also for $n=0$ which is in the \textit{different} universality class discussed in the previous section) are shown in Fig.~\ref{fig:all_scaling}.
 In all cases the scaling collapse is of similar quality to that in Fig.\ \ref{fig:S1_scaling}(b). 
  The corresponding values of $p_c$ and $\nu$ are listed in Table \ref{tab:exps}.  
For all the `physical' entropies ($n=1$, $n=2$, $n=\infty$) we find  consistent $p_c$ estimates for a given type of dynamics, and we find consistent   $\nu$ estimates for both types of dynamics. An uncertainty-weighted average of all six measured values at $n > 0$ gives 
\be
\nu = 2.03 \pm 0.05.
\ee
The functional form of the scaling functions at large positive and negative values of the scaling variable also appear compatible with what is implied by matching to linear growth of entanglement in the entangling phase and saturation in the disentangling phase.

For a given $n=1, 2$ or $\infty$, the  shape of the scaling functions looks slightly different for the two dynamical protocols (Floquet and random unitary). However, note that the  scaling function depends on the aspect ratio, which differs between the two cases (note that the aspect ratio  should be thought of not as $t/L$ but as $ut/L$, where $u$ is a nonuniversal speed that can differ for the two protocols).

Our scaling analysis above assumed that the characteristic lengthscale $\xi$ and the characteristic timescale $\tau$ diverge with the same power of $|p-p_c|$, rather than behaving as $\xi \sim |p-p_c|^{\nu_L}$ and $\tau\sim |p-p_c|^{\nu_t}$ with distinct $\nu_L$ and $\nu_t$, and a dynamical exponent  $z = \nu_t/\nu_L$ different from $1$.
Evidence for $z=1$ comes from comparing results for the entanglement in the limits $t \rightarrow \infty$ and $L \rightarrow \infty$.  We approximate these limits by extrapolating the data for random unitaries at fixed $L$ as a function of $t$, and then separately extrapolating the data at fixed $t$ as a function of $L$.  A separate scaling analysis was performed for each set of extrapolated data, and the resulting estimates for the critical exponents were $\nu_L = 1.99 \pm 0.20$ and $\nu_t = 2.00 \pm 0.37$.  These are consistent with each other and with the other results in Table \ref{tab:exps}.

\subsection{Two-point spatial correlations}

\begin{figure}[htb]
\begin{center}
\includegraphics[width=\linewidth]{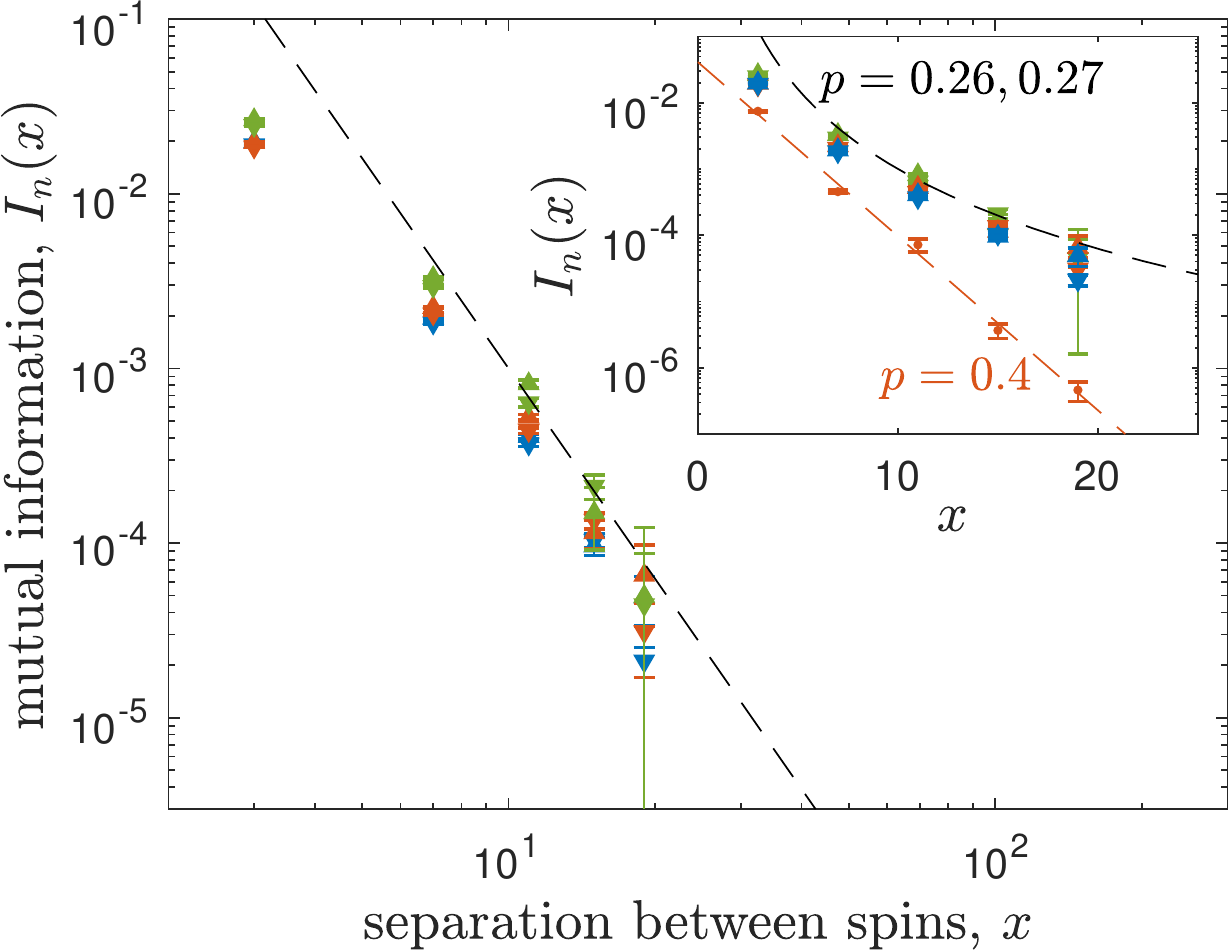}
\end{center}
\caption{Dependence of the mutual information $I_n$ between two spins on their separation $x$.  The main figure shows $I_n(x)$ for values of $p$ that are close to the critical value $p_c$.  The plotted data is taken from simulations of the random unitary dynamics, for which $p_c = 0.267 \pm 0.027$.  Blue, red, and green symbols correspond, respectively, to $n=1$, $n=2$, and $n\rightarrow \infty$, while upward- and downward-facing triangles correspond, respectively, to $p = 0.26$ and $0.27$. The dashed line shows the relation $I \propto 1/x^{4}$, as in the classical problem. The inset shows this same data on a semi-logarithmic scale, along with results for $I_1(x)$ at $p = 0.4$, which is away from the critical point.  These latter results are well fit by an exponential dependence (red dashed line). }
\label{fig:In}
\end{figure}

Equal-time correlations at the critical point can also be probed using the R\'{e}nyi mutual informations, as discussed in Sec.~\ref{sec:I0correlations}.\footnote{We note that for $n>1$ the R\'enyi entropies are not subadditive, and therefore $I_n(\mathbf{a},\mathbf{b})$ can take negative values.}  As in the classical case, we consider the quantity $I_n(\mathbf a, \mathbf b)$, defined by Eq.\ (\ref{eq:I0defn}), where the sets $\mathbf a$ and $\mathbf b$ constitute single spins separated by a spatial distance $x$.  (The quantity $I_2(x)$ is directly related to the spin-spin correlation function in a way that is described in footnote 5.)  In Fig.\ \ref{fig:In} we plot $I_n(x)$ close to the critical point of the random unitary dynamics for $n = 1,2$ and $n\rightarrow \infty$.  In order to minimize finite-size effects, we choose the system size to scale with $x$ at large $x$, setting $L=2(x+1)$ and $t=L$. The data corresponds to system sizes $L = 8, 16, 24, 32$, and $40$, and the corresponding separations are $x = 3, 7, 11, 15$, and $19$.  The data are consistent with power-law scaling at the critical point. 
For comparison, the black dashed line in Fig.\ \ref{fig:In} shows $I \propto x^{-4}$, as in classical percolation.  Away from the critical point, $I_n(x)$ shows exponential decay: see Fig.\ \ref{fig:In} inset, which shows data for $p=0.4$, and the discussion in Sec.\ \ref{sec:I0correlations}.

\section{Higher dimensions}
\label{sec:higherdimensions}

The mapping between $S_0$ and classical percolation implies that 
there is a transition in $S_0$ in any number of dimensions.
In this section we discuss the transition in higher dimensions, focusing for simplicity on $S_0$ only.  That is, we discuss the higher-dimensional version of our toy model.

\begin{figure}[t]
\begin{center}
\includegraphics[width=0.45\linewidth]{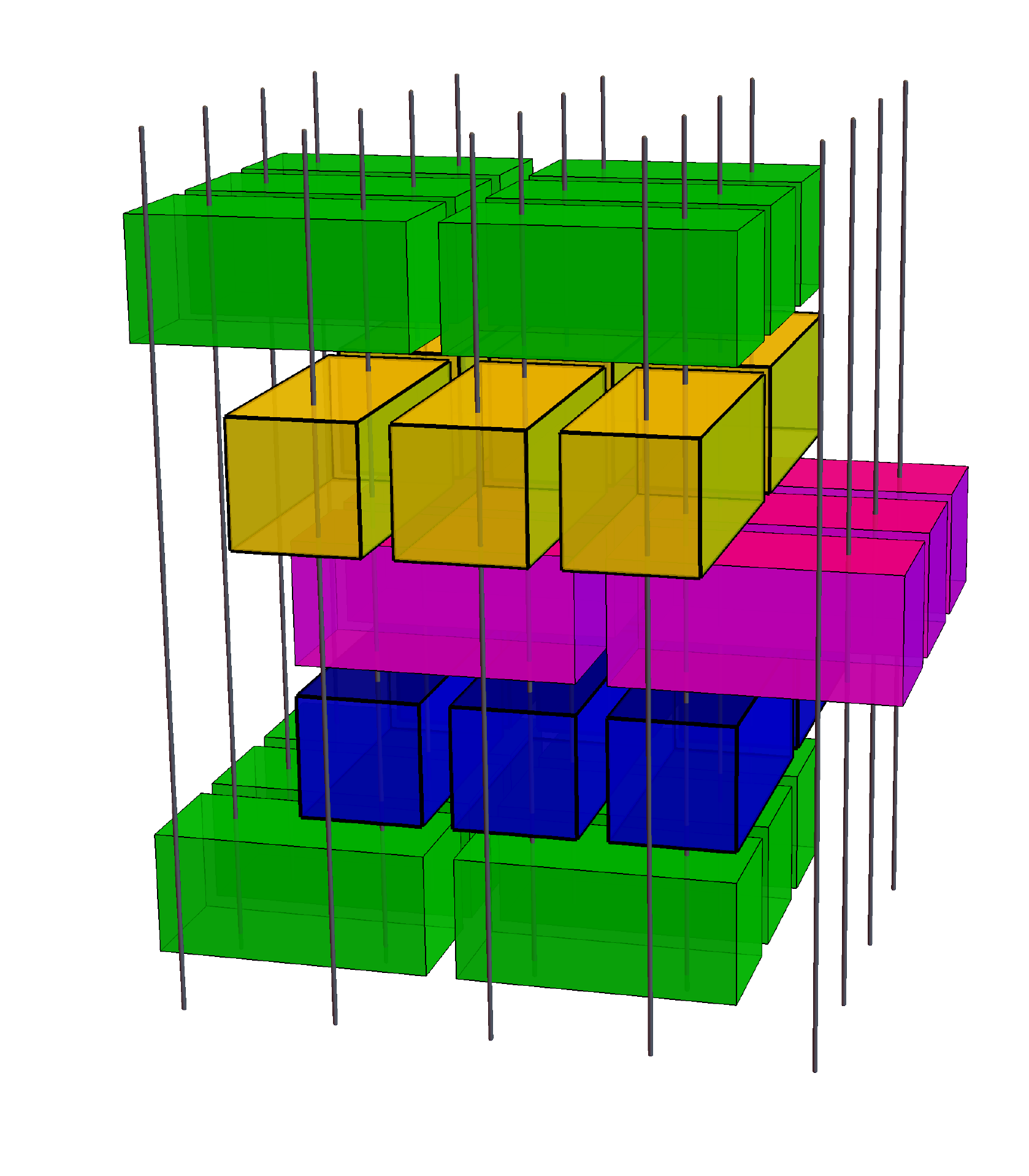}
\hspace{3mm}
\includegraphics[width=0.36\linewidth]{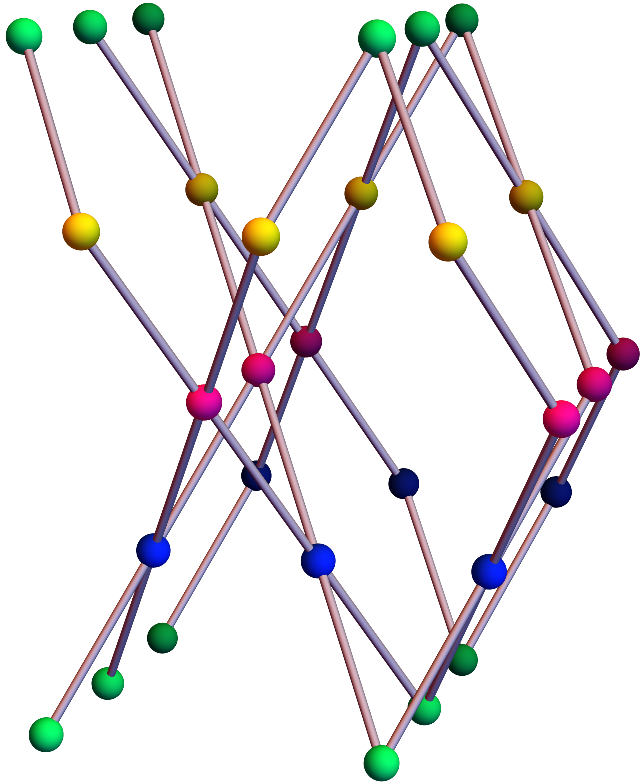}
\end{center}
\caption{
A 2+1D analogue of the correspondence in Fig.~\ref{fig:perclatticemapping}.
Left: a simple choice of circuit  geometry  in 2+1D.
Some unitaries (bricks) are hidden in the figure, so as to expose just six in each layer.
Right: each unitary corresponds to a node of the four-coordinated lattice shown.
Measurements break bonds of this lattice. The minimal cut (not shown) is a two-dimensional sheet whose cost is the number of unbroken bonds it bisects. }
\label{fig:3dlattice}
\end{figure}

The mapping of $S_0$ to the  optimal cost of a cut generalizes directly from the $1+1$D case.
For example, Fig.~\ref{fig:3dlattice} (Left) shows one choice of circuit geometry in 2+1D.
This leads to bond percolation problem on the lattice shown in the right panel.
In such higher dimensional situations, the minimal cut is a membrane, whose cost is equal to the number of unbroken bonds that pierce it.
To compute the zeroth  entanglement entropy $S_0(A)$ of a region $A$, one must find the minimal membrane separating the legs at the top in $A$ from those in $\overline{A}$.

As in the 1+1D case, there is a phase transition between an entangling phase at $p<p_c$ and a disentangling phase at $p>p_c$.
The disentangling phase has only area law entanglement in the steady state,
while the entangling phase has linear--in--time entanglement growth
and volume-law entanglement in the steady state.
A membrane picture for the case without measurements was introduced in  \cite{nahum_quantum_2017},
and applies similarly here.
The coefficient of the volume law vanishes as $p\rightarrow p_c$ from below, as in 1+1D.

The transition is at the bond percolation threshold for the lattice shown.\footnote{For various four-coordinated lattices in 3D, the bond percolation threshold is at a fraction $\sim 0.6$ of broken bonds \cite{vyssotsky1961critical,sykes1964critical,gaunt1983series,xu2014simultaneous}.}
Recall that in our notation $p$ is that the probability that a bond is \textit{broken}. When $p>p_c$, the unbroken bonds do not percolate. It is then easy to see that the cost per unit area vanishes for a large minimal membrane, leading to area law entanglement in the steady state.  When $p<p_c$, the unbroken bonds percolate. As a result the membrane must cross a number of unbroken bonds that scales with its surface area. This leads to the properties of the entangling phase mentioned above.

One difference from 1+1D is that this mapping indicates that the critical state exactly at $p_c$ is likely to show an area law for $S_0$ in spatial dimensions ${d>1}$.\footnote{This follows from standard domain wall scaling arguments \cite{fisher1997stability} if we assume that the cost $C(\ell)$ of a section of minimal membrane of scale $\ell$ grows more slowly than $\ell^{d-1}$ (given that this section is free to optimize its configuration at this scale). A plausible expectation is ${C(\ell)\sim 1}$ for all $d$, but this should be checked.}
This is reminiscent of \textit{ground} states of higher-dimensional critical systems described by conformal field theories, which show area law entanglement, in contrast to the 1+1D case where they show logarithmic entanglement \cite{calabrese2009entanglement}. Here the dynamical transition is between area law and volume law phases, so it is unlike a ground state phase transition, which is between area law phases.

\section{Discussion}

\subsection{Summary}

This paper has presented a new dynamical phase transition in the structure of evolving quantum wavefunctions.  
The generic phase diagram as a function of the measurement rate $p$ per degree of freedom is proposed in Fig.~\ref{fig:phasediagram}, with an `entangling' phase at low $p$ and a `disentangling' phase at large $p$.  The existence of a critical measurement rate $p_c$ means that one can induce a scale-invariant entanglement structure 
simply by tuning the frequency of measurements, without any fine-tuning of the Hamiltonian.

It should be emphasized, however, that in order to observe the distinction between the two phases it is essential to 
consider histories of the system with particular outcomes of the individual measurements, rather than simply averaging the density matrix over all possible measurement outcomes. The transition is not apparent in this averaged density matrix,  which for any measurement rate $p>0$ will generically just tend to the infinite temperature density matrix.
We comment on possible practical implications of our results in Sec.~\ref{sec:implications} below.

It is perhaps instructive to comment on the failure of a naive argument that would seemingly imply area-law entanglement at all $p>0$. Consider the 1+1D case.
In the absence of measurements, entanglement is produced at a nonzero rate.
It is tempting to imagine that when the entanglement $S$ between two (say, semi-infinite) subsystems is large, there is an associated lengthscale of order $\ell\sim S$, such that measuring a spin within a distance $\sim \ell$ of the division between the subsystems eliminates an $\mathcal{O}(1)$ amount of entanglement, while measuring  spins outside this region has a negligible effect. 
This would suggest that the steady-state entanglement is determined by balancing the $\mathcal{O}(1)$ production rate against a decay rate of order $p \ell \sim p S$ given by the total rate for measurements in this region. This balance would yield area law entanglement (of order $1/p$) even for small $p$.

The failure of this argument can be seen by thinking about the entanglement using the minimal cut picture. Consider, for example, the effect of a single measurement within the $\ell$--sized region. When the system size $L \rightarrow \infty$ and the time $t$ is finite, this measurement typically this has a very small effect on the entanglement, because it results in a broken bond (at the final time) that is far from the  path of the optimal cut, which is the shortest path to the initial-time boundary. 
The situation is different at asymptotically late times in a finite system; the minimal cut then goes sideways to the spatial boundary, and may take advantage of the broken bond. In this regime, measurements (in the smaller subsystem) do have an $\mathcal{O}(1)$ effect on the entanglement.

\subsection{Implications for simulations of quantum systems}
\label{sec:implications}

Our results may have useful implications for numerical simulations of many-body quantum systems. 
In particular, the entangling-disentangling transition is likely to be relevant to situations where we need to simulate evolutions (of pure quantum states) that involve measurements or effective measurements.

First, imagine that we wish to find the state at time $t$, denoted $\ket{\Psi^{(o_1,\ldots,o_N)}}$, given  knowledge of the outcomes $o_1, o_2 \ldots, o_N$ of the sequence of earlier measurements.
We assume that the Hamiltonian and initial state are also known, and that the number of measurements is extensive in both space and time.
In 1+1D, matrix product methods may be used for the evolution (in principle, similar considerations arise in higher dimensions). The computational difficulty of these methods depends on the amount of entanglement across cuts that divide the system into two parts \cite{vidal2003efficient,verstraete2006matrix}. Therefore, which side of the entangling-disentangling transition the system is on becomes crucial.
In the entangling phase (fewer measurements), a matrix product representation requires a bond dimension that scales exponentially with the system size, and  simulations quickly become unfeasible with growing system size.  On the other hand, in the disentangling phase, where the von Neumann entropy saturates to a finite value, we can expect rapid convergence as a function of bond dimension, so that the dynamics is computationally tractable. 

A natural context for such problems is the simulation of open quantum systems \cite{breuer2002theory}.
The dynamics of open systems is effectively nonunitary due to interaction with the environment. 
For example, systems of cold atoms or molecules may exchange photons with the environment, or may be subject to environmental noise that must be averaged over.
Formally, one can think of the environment as measuring the internal quantum states of the particles, except that the outcome of these measurements are not known and should be averaged over, yielding a mixed state.
A direct calculation of the time-dependence of quantum expectation values would require one to numerically evolve the full density matrix, 
which can be prohibitively difficult since the number of elements in the density matrix scales as the square of the Hilbert space dimension.

Because of this limitation, an important tool for calculations in open systems is the method of ``quantum trajectories", or ``quantum jumps" \cite{dum1992monte, dalibard1992wavefunction, molmer1993monte, Gleyzes2007quantum, garrahan2010thermodynamics, bauer2011convergence, lesanovsky2013characterization,  Murch2013observing,  tilloy2015spikes, bauer2015computing, bauer2017stochastic}
(see Refs.\ \onlinecite{plenio1998quantum,daley2014quantum} for reviews).  In this method, one calculates the evolution of one single pure state at a time, choosing at each instance of ``measurement'' a single  random outcome.
The time-dependence of an observable's expectation value may then be found by computing the expectation value in each simulated pure state, and then averaging over many such random trajectories. 
In the conceptually simplest case the effective dynamics of the pure state involves only unitary evolution and measurements. (More generally, additional non-Hermitian terms are required in the Hamiltonian, see e.g.\ Refs.\  \onlinecite{plenio1998quantum,daley2014quantum};
 the effect of such terms on the entanglement structure deserves further study.)

As noted above, the computational difficulty of such calculations is determined by the amount of entanglement in individual trajectories.\footnote{In the entangling phase, there may be alternative methods for efficiently simulating the state, taking advantage of the fact  that \textit{local} correlations are weak in typical volume-law states, despite the large entanglement (see Refs.~\cite{znidaric2016diffusive, leviatan2017quantum,white2018quantum,hallam2018lyapunov} for discussions of the unitary case); 
for example a matrix-product-operator representation of the reduced density matrix could be more efficient.
Therefore the vicinity of the phase transition could in fact be the most difficult to probe. 
In some circumstances nontypical realizations with large fluctuations in the entanglement could also be important. Large fluctuations of local observables in trajectories have been addressed in Refs.~\cite{garrahan2010thermodynamics,ates2012dynamical,hickey2013time}.}
We have shown in this paper that there is a sharp phase transition between different regimes, implying  a well-defined easy--to--hard transition for such numerical simulations as a function of the rate of dissipation to the environment. Interestingly, the logarithmic growth of the entanglement at the critical point implies a power-law scaling of computational difficulty;
this power-law scaling is also advantageous for numerical investigations of the entanglement phase transition.

This phenomenology of different regimes of entanglement and corresponding computational difficulty is consistent with numerical simulations presented in Ref.\ \onlinecite{Bonnes2014superoperators}.  The authors of Ref.\ \onlinecite{Bonnes2014superoperators} studied the time evolution of the entanglement in quantum trajectories of the Bose-Hubbard model at different rates of dissipation to the environment. When the dissipation was low, the entanglement was seen to grow as a function of time before saturating at a system-size-dependent value. At high dissipation, on the other hand, the entanglement quickly saturated to a small, finite value.  The authors also found that the quantum trajectory method becomes more efficient computationally when the dissipation is high.

A scaling analysis of how the entangling-disentangling transition affects computational hardness might be interesting. This may involve studying the critical scaling of the R\'{e}nyi entropies $S_n$ with $0<n<1$, which are important in the analysis of convergence of matrix product state algorithms
\cite{verstraete2006matrix}.

\subsection{Universality classes}
\label{sec:universalityclasses}

In this paper we have studied a toy model for the entangling-disentangling transition as well as what we propose is the generic version of this transition.
While similar in many respects, these problems are in different universality classes. In certain limits it should be possible to study a crossover between these universality classes.

In principle studying the entropies $S_n$ as a function of $n$ with $0 < n < 1$ might reveal such a crossover.
However, this does not give an obvious starting point for an analytical treatment. 

A more promising direction may be to attempt to expand around a limit of large local Hilbert space dimension.
We have focused our discussion on chains or lattices of spin$-1/2$.  
However, the random-unitary-dynamics-plus-measurement protocol we have discussed may be generalized to `spins' with local Hilbert space dimension $q\geq 2$.
If the minimal cut formula holds exactly in the limit $q\rightarrow\infty$ for random unitary dynamics plus measurements,
as it does for random unitary dynamics at infinite $q$  \cite{nahum_quantum_2017, zhou2018emergent} and for tensor networks with infinite bond dimension \cite{vasseur2018entanglement},
then taking $q\rightarrow \infty$ is an alternative way to obtain the effective classical optimization problem in Sec.~\ref{sec:zerothRenyi}, this time for all $S_n$ and not just $S_0$.\footnote{It would also be interesting to examine whether there are there special types of dynamics at \textit{finite} $q$ for which classical model is exact for all $n$. In purely unitary case there are examples of circuit dynamics for which $S_n$ is $n$ independent.}
Therefore in the regime $1 \ll q < \infty$ we expect to be able to probe a crossover between the toy model universality class of Sec.~\ref{sec:zerothRenyi} and the generic universality class of Sec.~\ref{sec:genericdynamicaltransition}.

When $q$ is large but finite there will then be a large crossover lengthscale $\xi_*$,
with the universal properties of the classical optimization problem visible at smaller scales and the those of the  `generic' universality class visible at larger scales.\footnote{If $1/q$ is increased from zero with $p$ fixed at the classical $p_c$ the flow will be to the disentangling phase. $p$ must also be slightly reduced from the classical $p_c$ to be located on the flow line to the generic critical point.}
Following the flow to scales $\gtrsim \xi_*$  is likely to be hard, but it should be possible to understand the initial instability.
Previous results for unitary circuits and tensor networks \cite{hayden2016holographic,nahum_operator_2017,chan_solution_2017, zhou2018emergent, vasseur2018entanglement} are suggestive of a possible mechanism. 
In simple limits, those mappings involve an effective statistical mechanics of domain walls that have both `energy' and configurational `entropy' (not to  be confused with the physical entanglement entropy). 
The crossover is likely to occur when configurational `entropy' becomes comparable with `energy' for these domain walls. 
For $q\gg 1$ energy dominates and is given by $\ln q$ times the length of the minimal cut. 
However `entropy' becomes significant at large scales $\ell$. 
Note that a segment of domain wall of minimal cost ($\ln q$) connecting two adjacent white clusters of scale $\sim \ell$  (Fig.~\ref{fig:perccriticalscaling}) has $\mathcal{O}(\ell^{3/4})$ choices for the connecting bond. This suggests that we should compare $\ln q$ with $\ln \ell^{3/4}$,  suggesting the crossover lengthscale $\xi_*\sim q^{4/3}$. This conjecture, however, demands a proper calculation.

A natural question, which we have not addressed here, is whether there is a field theory description of the renormalization group fixed point governing the entangling-disentangling transition. 
A starting point may be the replica trick, which has been used in unitary circuits and random tensor networks \cite{zhou2018emergent, vasseur2018entanglement} to construct mappings to (replica limits of) effective classical spin models in which the degrees of freedom are permutations \cite{hayden2016holographic}.\footnote{Classical mappings obtained
from purely unitary dynamics {($p=0$)} have crucial structure arising from the unitarity of the quantum dynamics   \cite{nahum_operator_2017, chan_solution_2017, zhou2018emergent}. For example, certain configurations have zero Boltzmann weight, simplifying the partition sum and allowing some exact results.
Including measurements will complicate this situation. 
First, for a given set of measurements the corresponding projection operators must be added to the circuit, violating unitarity. (See \cite{hayden2016holographic, vasseur2018entanglement} for other non-unitary tensor networks.) 
Second, the  measurement outcomes, which dictate the choice of projection operators, must be averaged over with a nontrivial joint probability distribution obtained from Born's rule, i.e. the components of the circuit are no longer independently random but acquire correlations.}

Notably, Ref.\ \onlinecite{vasseur2018entanglement} constructed a replica description of a `holographic' random tensor network state (projected entangled pair state)  and used it to argue that there is a phase transition between area law and volume law phases of the tensor network state.
While the transition itself is  hard to address as a result of the replica limit, the assumption of a single transition described by a conformal field theory leads to logarithmic scaling of the entanglement with subsystem size, and scaling forms analogous to Eq.~(\ref{eq:scalingformt}). The replica description also gives a formal explanation for why all the R\'{e}nyi entropies become simultaneously critical in the tensor network: they correspond to distinct correlation functions in the same conformal field theory \cite{vasseur2018entanglement}.

Above we found that the dynamical exponent for our dynamical transition is consistent with $z=1$, so it may be that the dynamical transition is governed by the same conformal field theory as the tensor network state  \cite{vasseur2018entanglement}, with one coordinate interpreted as time.
This is an interesting subject for future numerical simulations.

\subsection{Outlook}

The results we have presented here suggest a number of further directions that are worth exploring.  
Most directly, it would be useful to make a more detailed simulation study of the quantum problem.  
While we have pointed out that the quantum and classical problems are similar in the sense of having a scale-invariant critical point at a nonzero measurement rate, with qualitatively similar scaling forms, there are universal distinctions between them that are worth exploring.
One starting point is the more detailed behavior of the mutual information $I_n(x)$ as a function of separation $x$.

It would also be enlightening to test whether,
as suggested in Sec.~\ref{sec:models},
the  universality class of Sec.~\ref{sec:genericdynamicaltransition} 
applies even for models without any randomness in the choice of when and where to measure. 
We could obtain a convenient circuit model by fixing the locations of measurements and using the strength of the interaction in a Floquet unitary to drive the transition.

The classical problem may also be interesting to explore further, particularly in higher dimensions.  In 2+1D, for example, finding the  entanglement $S_0$ amounts to searching for a minimal surface, which is an interesting statistical mechanical problem. A numerical study might give insight into the entangling-disentangling transition in higher dimensions, where quantum numerics are challenging.

Finally, it will be interesting in the future to examine more subtle features of the evolving states. 
For example, in what respects is the entangling phase similar to unitary entangling dynamics and in what respects is it dissimilar? (Note, for example, that while unitary dynamics does not decrease entropy, the entangling dynamics at $0<p<p_c$ can reduce the entropy of a volume-law state of a finite system if its initial entropy density is higher than the steady state value.) How is the membrane picture for entanglement growth \cite{nahum_quantum_2017,jonay2018coarse, mezei2018membrane} modified by weak breaking of unitarity?
Finally, one could also ask how the `complexity' of the evolving state (the minimal number of local operations required to generate it from a product state \cite{knill1995approximation, nielsen2005geometric, nielsen2006quantum, susskind2016computational, stanford2014complexity, roberts2015localized, brown2016holographic, brown2017quantum, brown2018second, roberts2017chaos, cotler2017chaos, liu2018generalized}) behaves as the entangling-disentangling transition is traversed.   

\ 

\noindent
\textit{Note added: } We would like to draw the reader's attention to two related parallel works --- by Chan, Nandkishore, Pretko, and Smith \cite{CNPS}; and by Li, Chen, and Fisher \cite{FisherZeno} --- which appeared on the arXiv simultaneously with our own.

\acknowledgements

We thank D.~Bernard, E.~Bettelheim, A.~De Luca, J.~Dubail, J.~Garrahan, S.~Gopalakrishnan, M.~Gullans, J.~Haah, D.~Kovrizhin, J.~Pixley, S.~Roy, R.~Vasseur, and H.~Wilming for useful discussions or correspondence. 
BS was supported as part of the MIT Center for Excitonics, an Energy Frontier Research Center funded by the U.S. Department of Energy, Office of Science, Basic Energy Sciences under Award No.~DE-SC0001088.
JR acknowledges a fellowship from the Gordon and Betty Moore Foundation under the EPiQs initiative (grant No.~GBMF4303).
AN acknowledges EPSRC Grant No.~EP/N028678/1.

\appendix

\section{Scaling analysis}
\label{sec:scalingdetails}

In Sec.\ \ref{sec:genericdynamicaltransition}, we found estimates for the critical measurement rate $p_c$ and the correlation length exponent $\nu$ by searching for scaling collapse among curves $S_n(p, L) - S(p_c, L)$ as a function of the single variable $(p-p_c)L^{1/\nu}$.  Our algorithm for performing this search is as follows.

For a given value of $p_c$ and $\nu$, one can define an objective function $R(p_c, \nu)$, which should be minimized by the search procedure, as follows.  First, we estimate the value $S(p_c, L)$ for each system size $L$ by using linear interpolation between the closest points on either side of $p_c$.  We then calculate the scaling variable $x = (p - p_c)L^{1/\nu}$ for each value of $p$ and $L$ in the dataset.  The result is a family of curves $y_L(x)$, one curve for each system size, where $y_L = S(p, L) - S(p_c, L)$.  The objective function $R$ is then defined as the sum of the mean-squared-deviations of each curve from their common mean, summed over all unique points $x_i$ in the data set.  That is,
\be
R = \sum_{i, L} \left[ y_L(x_i) - \bar{y}(x_i) \right]^2.
\ee
Here, $y_L(x_i)$ indicates the value of $y_L$ at the point $x_i$; if this value has not been calculated explicitly, then it is estimated by linear interpolation.  $\bar{y}(x)$ is the average of $y_L(x)$ over all system sizes $L$.  If the point $x_i$ lies outside the range of values of $x$ that have been simulated for some curve $y_L(x)$, then this term is not included in the sum.  That is, when calculating the dispersion between curves at some value of $x$, only those curves for which there is data at the given $x$ are taken into account.

Given a set of simulation data and an estimate for $p_c$ and $\nu$, one can evaluate the objective function $R(p_c, \nu)$ numerically.  We then search numerically for the values of $p_c$ and $\nu$ that minimize the objective function.  This is done using simple gradient descent, and implemented in Matlab.  Care was taken to begin the search at different initial guesses for $p_c \in (0, 1)$ and $\nu \in (0, 10)$ to ensure that the solution found for each data set was globally optimal.

In order to estimate the uncertainty in our results for $p_c$ and $\nu$, we examine how our results change when the amount of data included in the scaling analysis is intentionally reduced.  In particular, we iteratively run the scaling analysis for a variety of reduced data sets, in which all data corresponding to system sizes larger than some value $L' \leq L_\text{max}$ has been removed.  (Here $L_\text{max}$ denotes the maximum system size in the data.)  The value of $L'$ is varied from $L_\text{max}/2$ to $L_\text{max}$. A rough estimate for $p_c$ or $\nu$ in the limit of $L' \rightarrow \infty$ can be made by making a linear fit of $p_c$ or $\nu$ as a function of $1/L'$ and taking the value of the $y$-intercept.  The uncertainty in $p_c$ or $\nu$ is estimated as the difference between the value of $p_c$ or $\nu$ at $L' = L_\text{max}/2$ and the extrapolated value at $L' \rightarrow \infty$.

The results of our search algorithm, and the corresponding uncertainties, are listed in Table \ref{tab:exps}.

%
%

\section{Simulation methods}
\label{sec:numericalmethods}

\subsection{Classical percolation simulation}

We use the following deterministic algorithm in order to find the minimal cut $S_0$ through a classical network --- namely, the dual square lattice illustrated on the right-hand side of Fig.\ \ref{fig:perclatticemapping} --- starting at some origin site $o$ on the dual lattice.  First, we define the adjacency matrix $\mathbb{A}$ for sites on the dual lattice, whose elements $\mathbb{A}_{ij}$ are such that $\mathbb{A}_{ij} = 1$ if site $i$ can be reached from $j$ by a single step (regardless of whether that step traverses a broken or an unbroken bond), and $\mathbb{A}_{ij} = 0$ otherwise.  The matrix $\mathbb{A}$ is entirely defined by the topology of the lattice and has no randomness. We also define the connectivity matrix $\mathbb{C}$, which depends on the times and locations of the measurements, and is defined so that $\mathbb{C}_{ij} = 1$ when the bond between $i$ and $j$ is broken and $\mathbb{C}_{ij} = 0$ otherwise.  Whether a site $j$ can be reached from a site $i$ without passing across an unbroken bond is determined by the ``wetting matrix'' $\mathbb{W}$, defined by
\be
\mathbb{W}_{ij} =
\begin{cases}
      1 & \textrm{if } \displaystyle{\lim_{n \to \infty}} (\mathbb{C}^n)_{ij} > 0 \\
      0 & \textrm{ otherwise}
\end{cases}
\ee
In practice, one need only calculate the $N$th matrix power of connectivity matrix, $\mathbb{C}^N$, where $N \sim L t$ is the number of bonds in the network.

Let $\mathbf{v}$ be a vector indicating which sites are ``wetted'' by the percolation process.  The algorithm to find the minimal cut begins with $\mathbf{v}_o = 1$, where $o$ indicates the index of the origin site, and $\mathbf{v}_i = 0$ for all other sites $i \neq o$.  Let $g$ indicate the index of the goal site; when calculating $S_0$ for a group of spins that includes the beginning or end of the chain, $g$ represents either the lateral or bottom boundaries of the lattice, and is adjacent to all sites along those boundaries.  If $(\mathbb{W} \mathbf{v})_g > 0$, then the goal site is ``wetted" by the origin site without the need to cross any unbroken bonds, and $S_0 = 0$.  Otherwise, one can use the following iterative process to find $S_0$:

\begin{enumerate}
\item Replace $\mathbf{v}$ with a vector indicating the set of all wetted sites, $\mathbf{v} \rightarrow \mathbb{W} \mathbf{v}$.
\item Add to the set of wetted sites those adjacent sites that are not wetted, $\mathbf{v} \rightarrow \mathbb{A} \mathbf{v}$.
\item If $\mathbf{v}_g > 0$, then the goal site has been reached.  If not, return to step 1.
\end{enumerate}

The minimal cut cost $S_0$ is equal to the number of times that steps 1 and 2 must be repeated before $\mathbf{v}_g > 0$.

\subsection{Quantum simulation}
\label{sec:quantumsimulation}

To simulate the quantum state evolution we use {\it exact} time evolution of a matrix product state. The matrix product state is manipulated using the ITensor library~\cite{ITensor}. The unitary circuit is divided into two-site unitary operations acting on pairs of adjacent spins, as described in Fig.~\ref{fig:firstcircuitfig} and as described in Ref. \cite{nahum_quantum_2017}. The two-site gates are applied, first over all odd numbered bonds (where the first bond on the left is numbered 1), and then over the even bonds.

As described in Section \ref{sec:models}, we consider two evolution protocols. The first, random unitary dynamics, utilizes two site random Haar gates. The gates are produced by randomly drawing 4 random vectors from a Gaussian distribution over the complex numbers. Then using the Gram-Schmidt procedure we obtain the desired unitary. 

The `Floquet' dynamics, on the other hand, has no inherent randomness except for the locations and outcomes of the measurements. Here we apply the two site gate, Eq.\ (\ref{eq:unitarydefinition}), in the same order as in the random dynamics. 

After a unitary is applied to a pair of spins,  each of the two spins may be measured, each with probability $p$. If a spin is measured, the outcome probabilities $p_\uparrow = |\bra{ \uparrow } \Psi(t) \rangle|^2$ and $p_\downarrow = 1-p_\uparrow$ are determined.  The state is then projected into a well defined spin state by applying the operator $(1\pm \hat\sigma^z)/2$ with probability $p_\uparrow$ for the up state and $p_\downarrow$ for the down state, and then renormalized.

For system lengths of even bond length, the entropy is computed each time a full double layer of unitaries is applied (i.e.~after the unitaries are applied to the even numbered bonds). On the other hand, for systems with an odd number of bonds, we compute the entropy after a {\it single } layer of unitaries is applied to the odd numbered bonds. 

To compute $S_0$ we use the fact that ITensor naturally minimizes the number of eigenvalues by discarding any redundant bond dimension. No truncation threshold is set for the number of eigenvalues or their size (thus they are cutoff by the numerical precision). For that reason, a certain number of spurious eigenvalues, with value of order of the numerical precision, are in general retained, producing a systematic (positive) error in the calculation of $S_0$.  In practice, however, we find that this systematic shift of the quantum simulation relative to the classical simulation is never larger than a few percent at any value of the simulation parameters that we examined.  Extending the simulation to much longer times or system sizes may increase the numerical inaccuracy of $S_0$.  We note, however, that the higher-order R\'{e}nyi entropies do not suffer from this same source of inaccuracy, since small eigenvalues contribute very little to them.

The typical number of realizations used to compute the entropy is 10,000. However, due to run-time constraints, we have often used fewer realizations for longer system sizes and small values of $p$. For example, for $t=24$ and $p=0.1$ we used only 500 realizations, and this is reflected in the larger error bars. Also note that, throughout the paper, error bars indicate one standard deviation of the result, as determined by the distribution of the result over all realizations, divided by the square root of the number of realizations.

To compute the mutual information $I_n(\mathbf a,\mathbf b)$ for spins  $\mathbf a$ and $\mathbf b$ we need to compute the entanglement entropy of two disconnected regions, which is not straightforwardly implemented in the ITensor library. We decompose the reduced density matrix of the two spins into a sum of Pauli operators 
\be 
\rho_{\mathbf a,\mathbf b} = \frac{1}{4} \sum_{\a,\b = 1,\ldots,4} w_{\a,\b}\; \s_{\mathbf a}^\a \otimes \s_{\mathbf b}^\b
\ee
where the coefficients are given by  
\be
 w_{\a,\b} = \mathrm{Tr} \left[\s_{\mathbf a}^\a \otimes \s_{\mathbf b}^\b \;\rho\right]  = \langle  \s_i^\a \otimes \s_j^\b  \rangle 
\ee
and where $\rho$ is the total density matrix. We note that here a much greater number of realizations is required ($\sim 100,000$). This is because the mean value of the mutual information is controlled by large values, which occur rarely.

\bibliographystyle{unsrt}
\bibliography{unitaryprojectorref}

\end{document}